\documentclass[preprint]{aastex}
\usepackage{graphicx}
\shorttitle{High-Contrast Imaging for Intermediate-Mass Giants}
\shortauthors{Ryu et al.}

\newcommand{\ms}{\mbox{m\,s$^{-1}$}}

\newcommand{\Msun}{\mbox{M$_{\odot}$}}

\newcommand{\Mjup}{\mbox{M$_{\rm Jup}$}}

\begin{document}
\title{High-Contrast Imaging of Intermediate-Mass Giants with Long-Term Radial Velocity Trends}

\author{Tsuguru Ryu\altaffilmark{1,2} , Bun'ei Sato\altaffilmark{3}, Masayuki Kuzuhara\altaffilmark{3}, Norio Narita\altaffilmark{1,2,4}, Yasuhiro H. Takahashi\altaffilmark{5}, Taichi Uyama\altaffilmark{5},
Tomoyuki Kudo\altaffilmark{6}, 
Nobuhiko Kusakabe\altaffilmark{4}, 
Jun Hashimoto\altaffilmark{4}, 
Masashi Omiya\altaffilmark{2},
Hiroki Harakawa\altaffilmark{2},
Lyu Abe\altaffilmark{7}, 
Hiroyasu Ando\altaffilmark{2},
Wolfgang Brandner\altaffilmark{8},
Timothy D. Brandt\altaffilmark{9}, \\
Joseph C. Carson\altaffilmark{10,8}, 
Thayne Currie\altaffilmark{6},
Sebastian Egner\altaffilmark{6}, 
Markus Feldt\altaffilmark{8}, 
Miwa Goto\altaffilmark{11}, \\
Carol A. Grady\altaffilmark{12,13,14}, 
Olivier Guyon\altaffilmark{6},
Yutaka Hayano\altaffilmark{6}, 
Masahiko Hayashi\altaffilmark{2}, 
Saeko S. Hayashi\altaffilmark{6},\\
Krzysztof G. He\l{}miniak\altaffilmark{6}
Thomas Henning\altaffilmark{8}, 
Klaus W. Hodapp\altaffilmark{15}, 
Shigeru Ida\altaffilmark{3},
Miki Ishii\altaffilmark{2}, \\
Yoichi Itoh\altaffilmark{16},
Masanori Iye\altaffilmark{2}, 
Hideyuki Izumiura\altaffilmark{1,17}
Markus Janson\altaffilmark{18}, 
Eiji Kambe\altaffilmark{17},\\
Ryo Kandori\altaffilmark{2}, 
Gillian R. Knapp\altaffilmark{18}, 
Eiichiro Kokubo\altaffilmark{1,2}
Jungmi Kwon\altaffilmark{5}, 
Taro Matsuo\altaffilmark{19}, \\
Satoshi Mayama\altaffilmark{1},
Michael W. McElwain\altaffilmark{20},
Kyle Mede,\altaffilmark{5},
Shoken Miyama\altaffilmark{21}, 
Jun-Ichi Morino\altaffilmark{2},\\
Amaya Moro-Martin\altaffilmark{22,23} 
Tetsuo Nishimura\altaffilmark{6}, 
Tae-Soo Pyo\altaffilmark{6},
Eugene Serabyn\altaffilmark{24}, \\
Takuya Suenaga\altaffilmark{1,2},
Hiroshi Suto\altaffilmark{2,4},
Ryuji Suzuki\altaffilmark{2}, 
Michihiro Takami\altaffilmark{25},
Naruhisa Takato\altaffilmark{6}, \\
Yoichi Takeda\altaffilmark{1,2},
Hiroshi Terada\altaffilmark{2},
Christian Thalmann\altaffilmark{26},
Edwin L. Turner\altaffilmark{19,27},\\
Makoto Watanabe\altaffilmark{28}, 
John Wisniewski\altaffilmark{29}, 
Toru Yamada\altaffilmark{30}, 
Michitoshi Yoshida\altaffilmark{31},
Hideki Takami\altaffilmark{2}, 
Tomonori Usuda\altaffilmark{2}, 
and
Motohide Tamura\altaffilmark{2,4,5}
}
\email{tsuguru.ryu@nao.ac.jp}

\altaffiltext{1}{SOKENDAI, The Graduate University for Advanced Studies, 2-21-1 Osawa, Mitaka, Tokyo 181-8588, Japan}
\altaffiltext{2}{National Astronomical Observatory of Japan, 2-21-1 Osawa, Mitaka, Tokyo 181-8588, Japan}
\altaffiltext{3}{Department of Earth and Planetary Sciences, Tokyo Institute of Technology, Ookayama, Meguro-ku, Tokyo 152-8551, Japan}
\altaffiltext{4}{Astrobiology Center, 2-21-1 Osawa, Mitaka, Tokyo, 181-8588, Japan}
\altaffiltext{5}{Department of Astronomy, The University of Tokyo, 7-3-1, Hongo, Bunkyo-ku, Tokyo, 113-0033, Japan}
\altaffiltext{6}{Subaru Telescope, National Astronomical Observatory of Japan, 650 North A�ohoku Place, Hilo, HI96720, USA}
\altaffiltext{7}{Laboratoire Lagrange (UMR 7293), Universite de Nice-Sophia Antipolis, CNRS, Observatoire de la Coted'azur,28 avenue Valrose, 06108 Nice Cedex 2, France}
\altaffiltext{8}{Max Planck Institute for Astronomy, K\"onigstuhl 17, 69117 Heidelberg, Germany}
\altaffiltext{9}{Astrophysics Department, Institute for Advanced Study, Princeton, NJ, USA}
\altaffiltext{10}{Department of Physics and Astronomy, College of Charleston, 58 Coming St., Charleston, SC 29424, USA}
\altaffiltext{11}{Universit\"ats-Sternwarte Mu\"nchen, Ludwig-Maximilians-Universit\"at, Scheinerstr. 1, 81679 Mu\"nchen,Germany}
\altaffiltext{12}{Exoplanets and Stellar Astrophysics Laboratory, Code 667, Goddard Space Flight Center, Greenbelt, MD
20771, USA}
\altaffiltext{13}{Eureka Scientific, 2452 Delmer, Suite 100, Oakland CA96002, USA}
\altaffiltext{14}{Goddard Center for Astrobiology}
\altaffiltext{15}{Institute for Astronomy, University of Hawaii, 640 N. A'ohoku Place, Hilo, HI 96720, USA}
\altaffiltext{16}{Nishi-Harima Astronomical Observatory, Center for Astronomy,University of Hyogo, 407-2, Nishigaichi, Sayo, Hyogo, 679-5313, Japan}
\altaffiltext{17}{Okayama Astrophysical Observatory, National
 Astronomical Observatory of Japan, Kamogata,
 Okayama 719-0232, Japan}
\altaffiltext{18}{Department of Astrophysical Science, Princeton University, Peyton Hall, Ivy Lane, Princeton, NJ08544, USA}
\altaffiltext{19}{Department of Astronomy, Kyoto University, Kitashirakawa-Oiwake-cho, Sakyo-ku, Kyoto, Kyoto 606-8502, Japan}
\altaffiltext{20}{Exoplanets and Stellar Astrophysics Laboratory, Code 667, Goddard Space Flight Center, Greenbelt, MD20771, USA}
\altaffiltext{21}{Hiroshima University, 1-3-2, Kagamiyama, Higashihiroshima, Hiroshima 739-8511, Japan}
\altaffiltext{22}{Space Telescope Science Institute, 3700 San Martin Drive, Baltimore, MD 21218, USA}
\altaffiltext{23}{Center for Astrophysical Sciences, Johns Hopkins University, Baltimore MD 21218, USA}
\altaffiltext{24}{Jet Propulsion Laboratory, California Institute of Technology, Pasadena, CA, 171-113, USA}
\altaffiltext{25}{Institute of Astronomy and Astrophysics, Academia Sinica, P.O. Box 23-141, Taipei 10617, Taiwan}
\altaffiltext{26}{ETH  Zurich,  Institute  for  Astronomy,  Wolfgang-Pauli-Strasse  27, 8093 Zurich, Switzerland}
\altaffiltext{27}{Kavli Institute for Physics and Mathematics of the Universe, The University of Tokyo, 5-1-5, Kashiwanoha, Kashiwa, Chiba 277-8568, Japan}
\altaffiltext{28}{Department of Cosmosciences, Hokkaido University, Kita-ku, Sapporo, Hokkaido 060-0810, Japan}
\altaffiltext{29}{ H. L. Dodge Department of Physics \& Astronomy, University of Oklahoma, 440 W Brooks St Norman, OK 73019, USA}
\altaffiltext{30}{Astronomical Institute, Tohoku University, Aoba-ku, Sendai, Miyagi 980-8578, Japan}
\altaffiltext{31}{Hiroshima Astrophysical Science Center, Hiroshima
University, Higashi-Hiroshima, Hiroshima 739-8526, Japan}

\begin{abstract}
A radial velocity (RV) survey for intermediate-mass giants has been operated for over a decade at Okayama Astrophysical Observatory (OAO). { The OAO survey has revealed that some giants show long-term linear RV accelerations (RV trends), indicating the presence of outer companions. 
Direct imaging observations can help clarify what objects generate these RV trends. }
We present the results of high-contrast imaging observations of six intermediate-mass giants with long-term RV trends using the Subaru Telescope and HiCIAO camera. 
We detected co-moving companions to $\gamma$ Hya B  ($0.61^{+0.12}_{-0.14}   M_\odot$), HD 5608 B ($0.10 \pm 0.01 M_\odot$), and HD 109272 B ($0.28 \pm 0.06  M_\odot$). { For the remaining targets($\iota$ Dra, 18 Del, and HD 14067) we exclude companions more massive than 30-60 $M_\mathrm{Jup}$ at projected separations of 1\arcsec--7\arcsec. We examine whether these directly imaged companions or unidentified long-period companions can account for the RV trends observed around the six giants. We find that the Kozai mechanism can explain the high eccentricity of the inner planets $\iota$ Dra b, HD 5608 b, and HD 14067 b.}
\end{abstract}


\section{Introduction}
The radial velocity (RV) technique has played a significant role in the search for exoplanets 
and has been used in the discovery of more than 500 planets in the last 20 years. However, the RV technique is less sensitive to wide-orbit planets with a semimajor axis larger than $\sim$10 AU. To confirm the existence of such planets, it is necessary to monitor the RV variation of the host star over an extremely long period, which is impractical. Hence, the occurrence rate of such wide-orbit planets remains poorly examined, even though it is a critical factor for testing planet formation/evolution theories such as core accretion (e.g., \citealt{pollack96}), gravitational disk instability (e.g., \citealt{durisen07}), and planet migration (e.g., \citealt{kley12}).

The long-term RV { acceleration (RV trends)} of a host star is useful information for uncovering possible planetary companions in wide orbits. If a companion exists beyond { $\sim$}10 AU from the host star, the companion generates an almost linear trend in the RV of the host star within a relatively short period. The slope of the trend depends on the mass and the semi-major axis of the RV trend generator (RVTG), and we can estimate its minimum mass based on the following relation:

\begin{equation}
\dot{v} \sim m_p \sin i \frac{\mathrm{G}}{a^2}
\end{equation}
where $m_p$ is the RVTG mass, $i$ is the orbital inclination, G is the gravitational constant, $a$ is the semi-major axis of the RVTG, and $\dot{v}$ is the RV trend. For example, {{ an RV trend of 10 m/s/yr}} corresponds to {{ 5$ M_\mathrm{Jup}$, the minimum mass of the RVTG at a semi-major axis of 10 AU.}} However, such an RV trend could just as easily be generated by a {{face-on or distant}} stellar companion or a brown-dwarf companion as a planetary one. A companion with $m_p \sin i \sim 0.5 M_\odot$ located at 100 AU also yields an
RV trend of 10 m/s/yr for the host star. Accordingly, the detection of the RV trend alone is not sufficient to identify the RVTG.

In contrast, direct imaging techniques are sensitive to such wide-orbit companions. Direct imaging technique can achieve a contrast better than $10^{-5}$ at a separation of 1\farcs0 from a central star (e.g., \citealt{suzuki10}) and thus can easily identify a stellar companion. Hence, this technique can help us to  clarify the true nature of an RVTG. Even non-detection of any companions is { useful for }constraining the range of the mass and semi-major axis of an RVTG by { a simultaneous analysis of the direct-imaging and RV-trend data.}

One study that employs this idea, namely, combining direct imaging {{and}} the RV-trend observations, is called TaRgetting bENchmark objects with Doppler Spectroscopy (TRENDS; e.g. \citealt{crepp12}), which attempts to detect companions around FGKM-type stars showing RV trends.
The study has discovered three low-mass stellar companions \citep{crepp12},
a tertiary stellar companion \citep{crepp13a}, 
a white dwarf companion \citep{crepp13b},
and a T dwarf \citep{crepp14}.
{ Also, they determined the giant planet occurrence rate around M-dwarf stars by exploring, via adaptive optics imaging, targets that exhibited an RV trend suggestive of an exoplanet companion \citep{Montet14}. }
Furthermore, this technique has revealed that stars hosting a hot-Jupiter tend to be accompanied by a stellar companion \citep{knutson14}.
These results clearly show that direct imaging observations can help us to explore and identify distant companions that generate RV trends in host stars.

  At Okayama Astrophysical Observatory (OAO), an RV survey targeting intermediate-mass giants (1.5--5 M$_\odot$) has been conducted for over a decade (e.g., \citealt{sato03}). 
\citet{sato08} found that there is a difference between orbits of planets around intermediate-mass stars and around lower-mass FGK stars. Most planets around intermediate-mass stars have a semi-major axis larger than 0.6 AU, while FGK stars have shorter-period planets. Hence, it was suggested that the orbital distribution of exoplanets around intermediate-mass stars is different from that around solar-type stars. In addition, the OAO survey detected long-term RV trends in several targets, which indicates the presence of distant companions around them. { The widest-orbit of planets or brown dwarfs so far discovered is 5 AU }\citep{sato13a}. Identifying the companions that generate the RV trend can improve our knowledge of exoplanet populations for intermediate-mass stars, which are not well understood compared to solar-type stars.

   To clarify the nature of the RVTGs around intermediate-mass stars observed in the OAO RV survey, we have performed direct-imaging observations as part of the Strategic Exploration of Exoplanets and Disks with Subaru (SEEDS; \citealt{tamura09}) project. 
SEEDS has discovered stellar companions around transiting planet systems \citep{narita10, narita12, takahashi13}, as well as planetary companions\cite[e.g.][]{kuzuhara13}.
While TRENDS targeted {{FGKM-type}} stars, our campaign has focused on intermediate-mass stars with RV trends and is therefore complementary to TRENDS.  
We imaged five stellar companions around these targets, and three companions are likely to be sources of RVTGs. In Section 2, we describe the RV and direct-imaging observations and the data reduction. In Section 3 we present the results of the direct imaging observations. In Section 4, we discuss the results and verify whether our imaged companions can generate the RV trends. We also discuss the orbit evolution of inner eccentric planets based on the Kozai mechanism. Finally, we summarize our results and discussion in Section 5.

\section{Observations \label{observation}}
{{ RV observations at OAO have identified five intermediate-mass giants with linear RV trends.  In order to clarify the objects that cause the RV trends, we observed the
 five giants via direct imaging.
Apart from the OAO survey, \citet{zechmeister08} and \citet{kane10} found 
a linear RV trend around another giant, $\iota$ Dra. The RVTG of $\iota$ Dra has 
been unclear, so we also carried out direct imaging observations of this giant.
 In total, we observed six intermediate-mass giants showing linear RV trends in the SEEDS campaign. 
Table \ref{targetproperty} shows the stellar properties of our six targets.}} Note that four of the targets have already known RV planets (Table \ref{innerplanets}). 
In this section, we describe our Doppler measurement observations, orbital fitting analysis to the RV data, and direct imaging observations.

\subsection{Doppler measurement observations}\label{dopplerobs}

We obtained RV data for the targets except for $\iota$ Dra with the 1.88-m
telescope and the HIgh Dispersion Echelle Spectrograph \citep[HIDES;][]{izumiura99}
at OAO between 2001 and 2014. We used an iodine absorption cell
\citep[I$_2$ cell;][]{kambe02} for precise RV measurements, which provides a fiducial
wavelength reference in a wavelength range of 5000--5800 ${\rm \AA}$.
We used the HIDES-slit mode setting with a slit width of the spectrograph of
200 $\mu$m ($0\farcs76$), which corresponds to a spectral resolution
($R=\lambda/\Delta\lambda$) of 67000 by about 3.3-pixel sampling.
Reduction of the echelle data (i.e., bias subtraction, flat-fielding,
scattered-light subtraction, and spectrum extraction) was performed
using the IRAF software package\footnote{IRAF is distributed by the National
Optical Astronomy Observatories, which is operated by the
Association of Universities for Research in Astronomy, Inc. under
cooperative agreement with the National Science Foundation,
USA.}.

\begin{figure}
\plotone{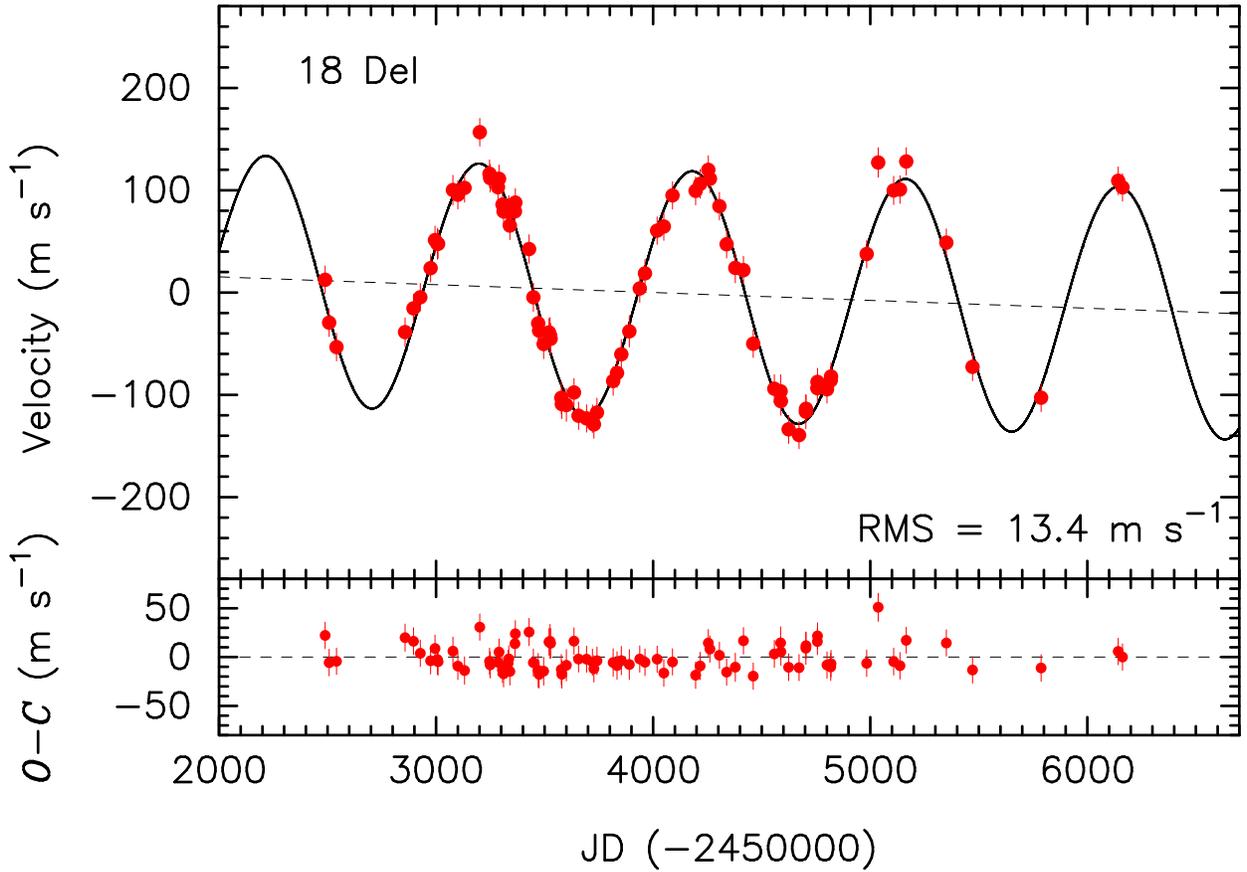}
\caption{Radial velocities of 18 Del observed at OAO. The
nearly circular Keplerian orbit with a linear velocity trend
($\dot{\gamma}= -2.8$ m s$^{-1}$ yr$^{-1}$) is shown
by the solid line. The error bar for each point includes the estimated
stellar jitter (12.9 \ms).
Bottom: Residuals to the orbital fit. The RMS to the fit is 13.4 \ms.}
\label{fig:18Del-RV}
\end{figure}

\begin{figure}
\epsscale{1.1}
\centering
\plotone{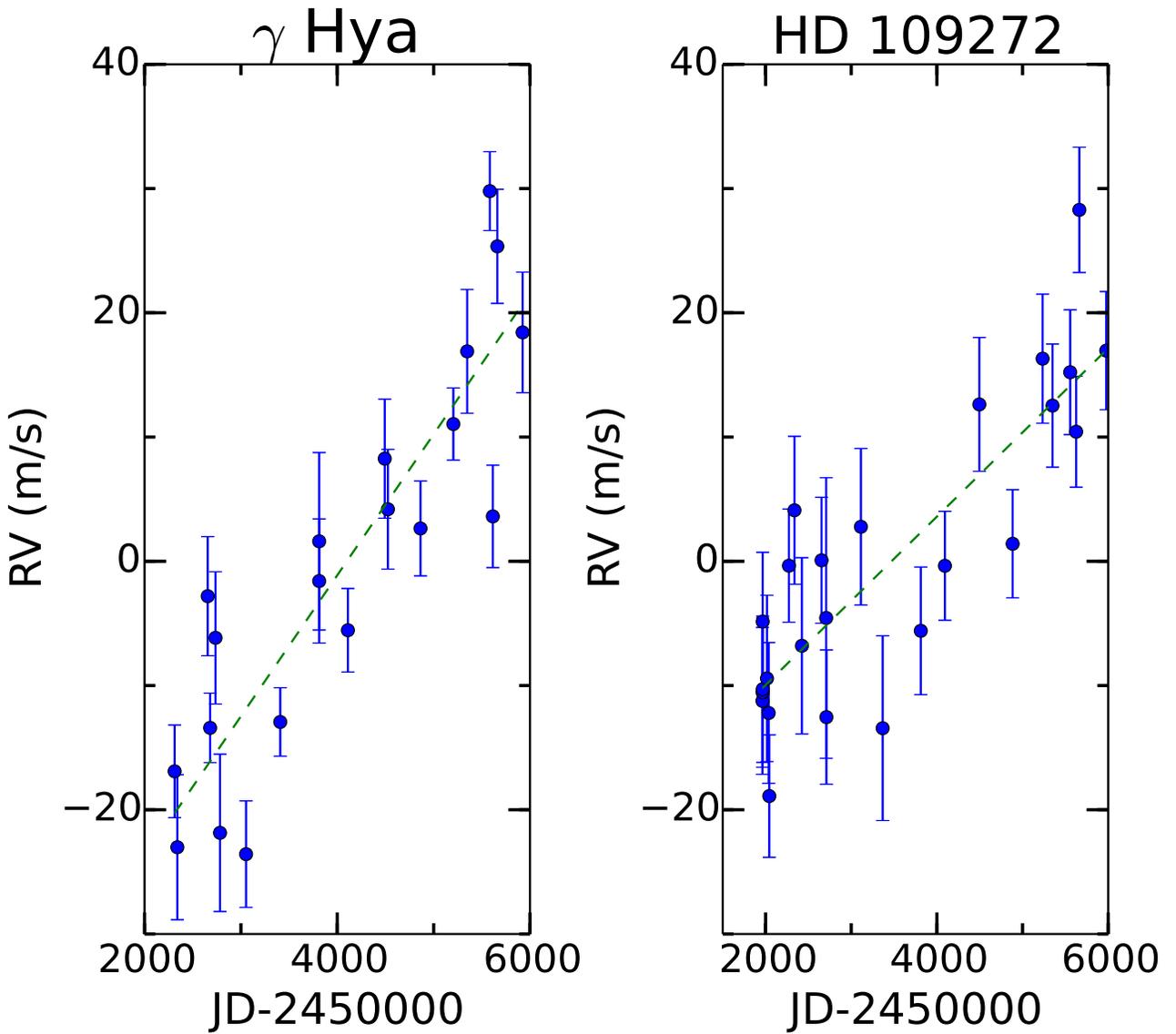}
\caption{Measured RV data of the two systems $\gamma$ Hya (left) and HD 109272 (right). The dashed lines show the best-fit linear trends. }
\label{rvplot}
\end{figure}

For precise RV analysis, we modeled I$_2$-superposed stellar spectra
(star+I$_2$) by the method detailed in \citet{sato:2002} and
\citet{sato12}, which is based on the method by
\citet{butler:1996} and \citet{val:95}.
In the method, a star+I$_2$ spectrum is modeled as a product of a
high resolution I$_2$ and a stellar template spectrum convolved
with a modeled instrumental profile (IP) of the spectrograph.
The stellar template spectrum is obtained by deconvolving a pure stellar
spectrum with an IP estimated from a B-star or flat spectrum
taken through a I$_2$ cell. We achieved a long-term RV precision
of about 4 m s$^{-1}$ over the entire span of the observations. 
The measurement error was derived from an ensemble of the velocities from each
of the $\sim$300 spectral segments (each $\sim$3$ {\rm \AA}$ long)
in every exposure. We show the derived RVs for 18 Del, $\gamma$ Hya, and HD 109272 
in Figure \ref{fig:18Del-RV} and Figure \ref{rvplot}, and have listed
them in Table \ref{tbl:18Del-RV}, Table \ref{rvvaluegamhya}, and Table \ref{rvvaluehd109272} together with the estimated uncertainties. The RVs for 18 Del were
updated and extended from those presented in \citet{sato08}.
The RVs for HD 5608 and HD 14067 presented in \citet{sato12}
and \citet{wang14}, respectively, were used for the
analysis in this paper.

\subsection{Orbital fitting}
After the first announcement of the discovery of a planet around 18 Del
by \citet{sato08}, we collected 31 more epochs of RV data for the star in five years
and updated its orbital parameters including a possible linear velocity trend
($\sim 4$ m s$^{-1}$ yr$^{-1}$) suggested in \citet{sato08}.
The updated orbital parameters and the uncertainties were derived using the Bayesian
Markov Chain Monte Carlo (MCMC) method (e.g., \citealt{ford:2005,gregory:2005,
ford:2007}), following the analysis in \citet{sato13b}.
An extra Gaussian noise factor {{representing stellar jitter}} for the data and a linear velocity trend were incorporated
as free parameters.
We generated 10 independent chains having $10^7$ points with
an acceptance rate of about 25\%, the first 10\% of which were discarded, and
confirmed that each parameter sufficiently converged based on the
Gelman--Rubbin statistic \citep{gelman:1992}.
We derived the median value of the merged posterior probability distribution function
(PDF) for each parameter and set the 1$\sigma$ uncertainty as the range
between 15.87\% and 84.13\% of the PDF.
We plot the derived Keplerian orbit together with the RV points and their
measurement errors including the jitter in Figure \ref{fig:18Del-RV}, and list
the orbital parameters and the uncertainties in Table \ref{tbl:18Del-orbit}.
We confirmed the linear velocity trend for the star to be $\dot{\gamma}= -2.8\pm0.7$
m s$^{-1}$ yr$^{-1}$ with 4$\sigma$ confidence.

For $\gamma$ Hya and HD 109272, we fit the linear velocity trends using the method of least squares. The trends of our targets are summarized in Table \ref{rvtrend_table}.

\subsection{Direct imaging observations}

Direct imaging observations were conducted from 2011 to 2014 as part of the SEEDS survey \citep{tamura09} using the High Contrast Instrument for the Subaru Next Generation Adaptive Optics (HiCIAO; \citealt{suzuki10}) on the 8.2-m Subaru Telescope. 
We used the adaptive optics system AO188 \citep{hayano08} together with HiCIAO, and the target stars themselves were used as the natural guide stars in our observations. In addition, we used an atmospheric dispersion corrector (ADC), which helped mitigate the drift of the stellar PSFs on the detector \citep{egner10}. Furthermore, the angular differential imaging (ADI) method \citep{marois06} was applied to our observations to improve the high-contrast performance. All six targets were first observed in 2011--2012 in the H-band($\sim$1.6 $\mu$m) and four targets that have companion candidates were followed up  in 2014 by employing J- ($\sim$1.2 $\mu$m), H-, and Ks- ($\sim$2.1 $\mu$m) band filters. 

In order to maximize the sensitivity of our observations, it is necessary to use an occulting mask or to saturate the central star's PSF. Such masked or saturated images (i.e., science images) can be taken over relatively long integration times and are used for the companion survey. However,  we require unsaturated and unmasked PSFs of the central stars (i.e., calibration images) for our measurements of the contrast limit, central star centroid. Therefore, we made unmasked observations with neutral density (ND) filters to avoid PSF saturation for each target, selecting ND filters of an appropriate transmission based on the central star's brightness: HiCIAO has ND filters with transmittances of 0.854, 0.063, and 0.016\%. We obtained the unsaturated and unmasked PSFs before and after observing the science images for each target. { The observation log is summarized in Table \ref{obslog}. }


Our data reduction procedure is as follows. First, we removed stripe patterns \citep[see][]{suzuki10} appearing on each observed frame and subsequently corrected hot and bad pixels. Then, hot-pixel masks were generated from dark frames obtained in each observing run. \linebreak
To create hot-pixel mask, we used L.A. Cosmic algorithm \citep{vandokkum01} for the data taken before 2012 September, while our originally-developed routine was applied to the data taken after 2014 April. 
{ Flat-fielding was performed following these procedures.}


{ Next, we correct the distortion of the images, since the distortion correction of our observed images is crucial to achieve reliable astrometry.  To measure the distortion map, we usually obtained images of the globular cluster M5 or M15 in each observing run.  The distortion map is made by comparing the stellar positions on the M5/M15 images taken by HiCIAO with those on the images taken by the Hubble Space Telescope/ACS whose distortion is well-corrected \citep{brandt13}.   Using the measured distortion map, we applied the geometric transformations to the post-flat-fielded frames.  This procedure fixes the plate scale of images to be 9.5 mas.}

{ We estimate the centroids of primary star in the distortion-corrected images using the unsaturated data whose distortions have been also corrected, and shift the positions of central star to the centers of arrays. Then, we assume that the stellar position on the detector did not drift. We also calculated the stellar frame-to-frame centroids by fitting the Moffat function to the masked PSFs and confirmed that the PSF drifts are less than 1 pixel (= 9.5 mas) during the observations.}

Next, we {{carried}} out ADI reductions { after subtracting} the central star's radial profile.  We used the locally optimized combination of images (LOCI) algorithm \citep{lafreniere07a}, which allows further improvement in our capability to detect faint companions. We evaluated the self-subtraction effect caused by the LOCI algorithm by embedding artificial PSFs. To determine a realistic detection limit, the final image was convolved with a circular aperture with a diameter equal to the PSF FWHM \citep{lafreniere07b}. Finally, we checked the achieved 5$\sigma$ contrast-ratio by calculating the standard deviation within 2-pixel wide rings, from the center to the outer region every 4.5 pixels.

\section{Results}\label{results}
We detected five companion candidates in the four systems $\gamma$ Hya, 18 Del, HD 5608, and HD 109272. Follow-up observations confirmed that three of the companion candidates in three systems, $\gamma$ Hya, HD 5608, and HD 109272, have a common proper motion. We converted the observed flux of the companion to its mass using the NextGen model \citep{hauschildt99a,hauschildt99b} or the Dusty model \citep{chabrier00}. The model { that} was most consistent with regard to the derived mass for all three bands (J-, H-, Ks-band) was adopted. {{The targets' ages, excepting $\iota$ Dra, were estimated in \citet{takeda08} by comparing the luminosities and effective temperatures with the theoretical stellar evolution model \citep{Lejeune2001}. We roughly estimated the age of $\iota$ Dra by comparing its luminosity and effective temperature with a theoretical model \citep{bressan12,chen15} }}

\subsection{Confirmed stellar companions}\label{companioncase}
\subsubsection{$\gamma$ Hya}
\begin{figure}
\plotone{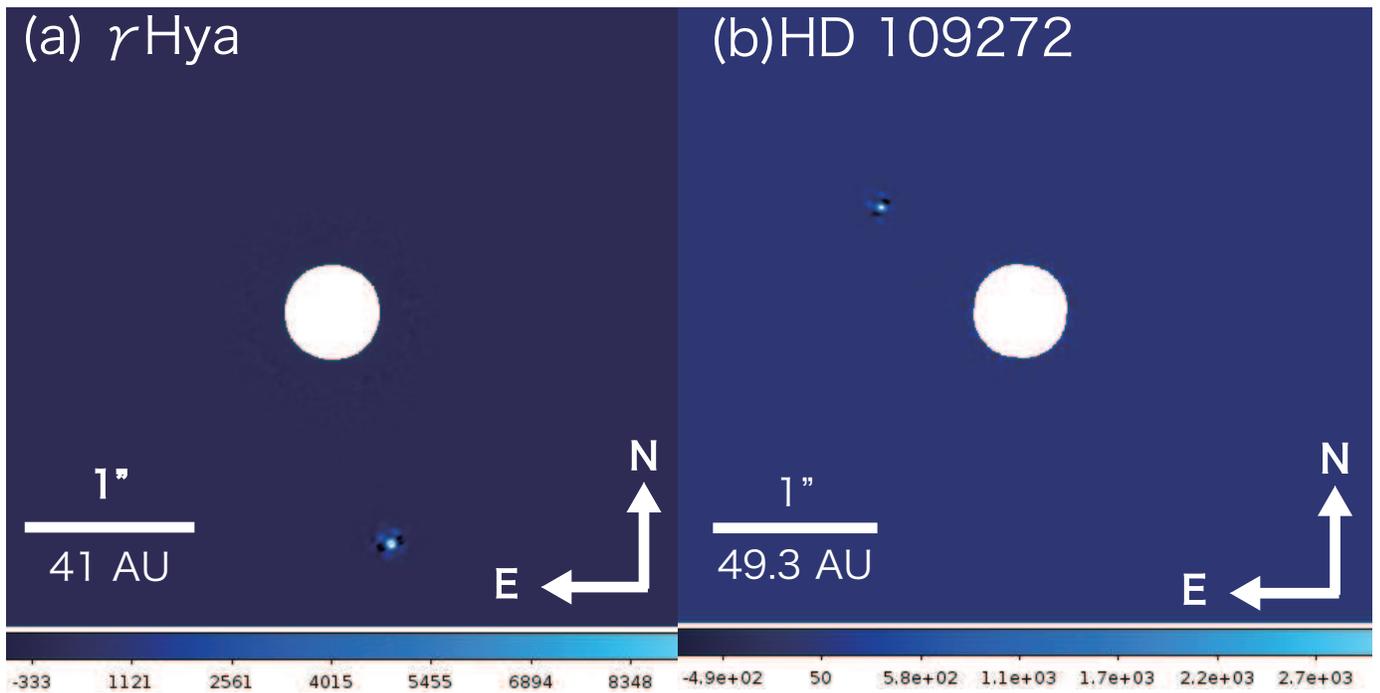}
\caption{Detected bright companions from HiCIAO observations. (a) Final image of $\gamma$ Hya in the H-band taken on 2012 May 13. North is up and east is left. The companion was detected at 1\farcs6 from $\gamma$ Hya.  (b) Final image of HD 109272 in the H-band taken on 2012 April 11. The companion candidate at 1\farcs2 can be seen in the figure. }
\label{hiciaoresult1}
\end{figure}

\begin{figure}
\plotone{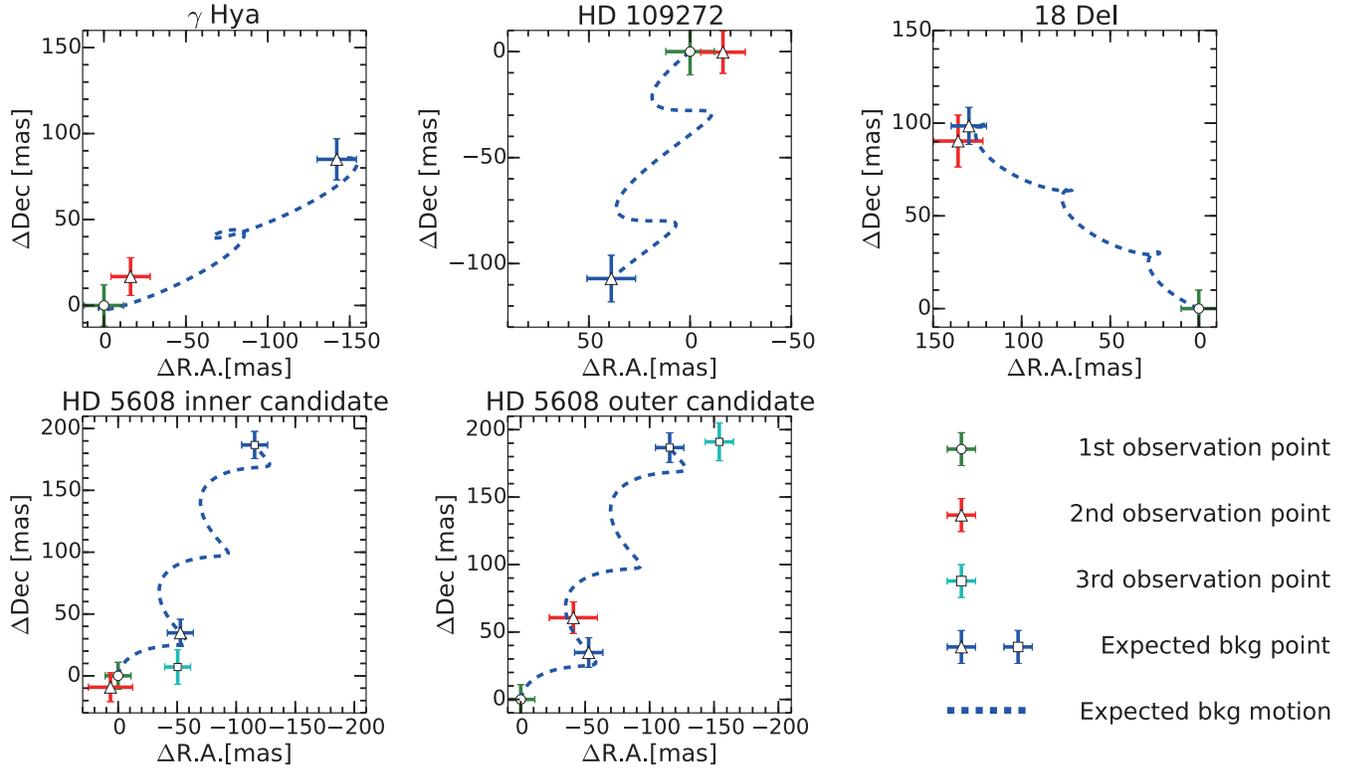}
\caption{Results of the common proper motion test for the companion candidates. The horizontal and vertical axes are relative distances from the first observation point. In each plot the circle with a green cross is the first observed position with error, the triangle with a red cross represents the second observed position, and the square with a cyan cross represents the third observed position. The blue dotted line represents the track of background star motion driven by the stellar parallax and the proper motion of each star. The blue crosses shows the positions of the observational data if the companion candidate is a background star. }
\label{cpmresult}
\end{figure}

We discovered a companion candidate with an H-band contrast of $\Delta${{H}} = 7.24 {{located}} 1\farcs6 from $\gamma$ Hya, as shown in Figure \ref{hiciaoresult1}. 
{{Two years after the first observation,}} a follow-up observation {{enabled}} us to confirm that the companion candidate, $\gamma$ Hya B, shares a common proper motion with the central star (Figure \ref{cpmresult}). Our astrometric and photometric results are shown in Table \ref{gamhyabresult}. Considering $\gamma$ Hya's age {{(0.37 Gyr, \citealt[][]{takeda08}) }} and consistency of the mass derived from J-, H-, and Ks-band photometry, we adopted the 400-Myr NextGen model to convert the measured photometry into mass. Mass of $\gamma$ Hya B is $0.61^{+0.12}_{-0.14}  ~ M_\odot$ { by averaging four independent mass estimates.}

\subsubsection{HD 5608}
\begin{figure}
\plotone{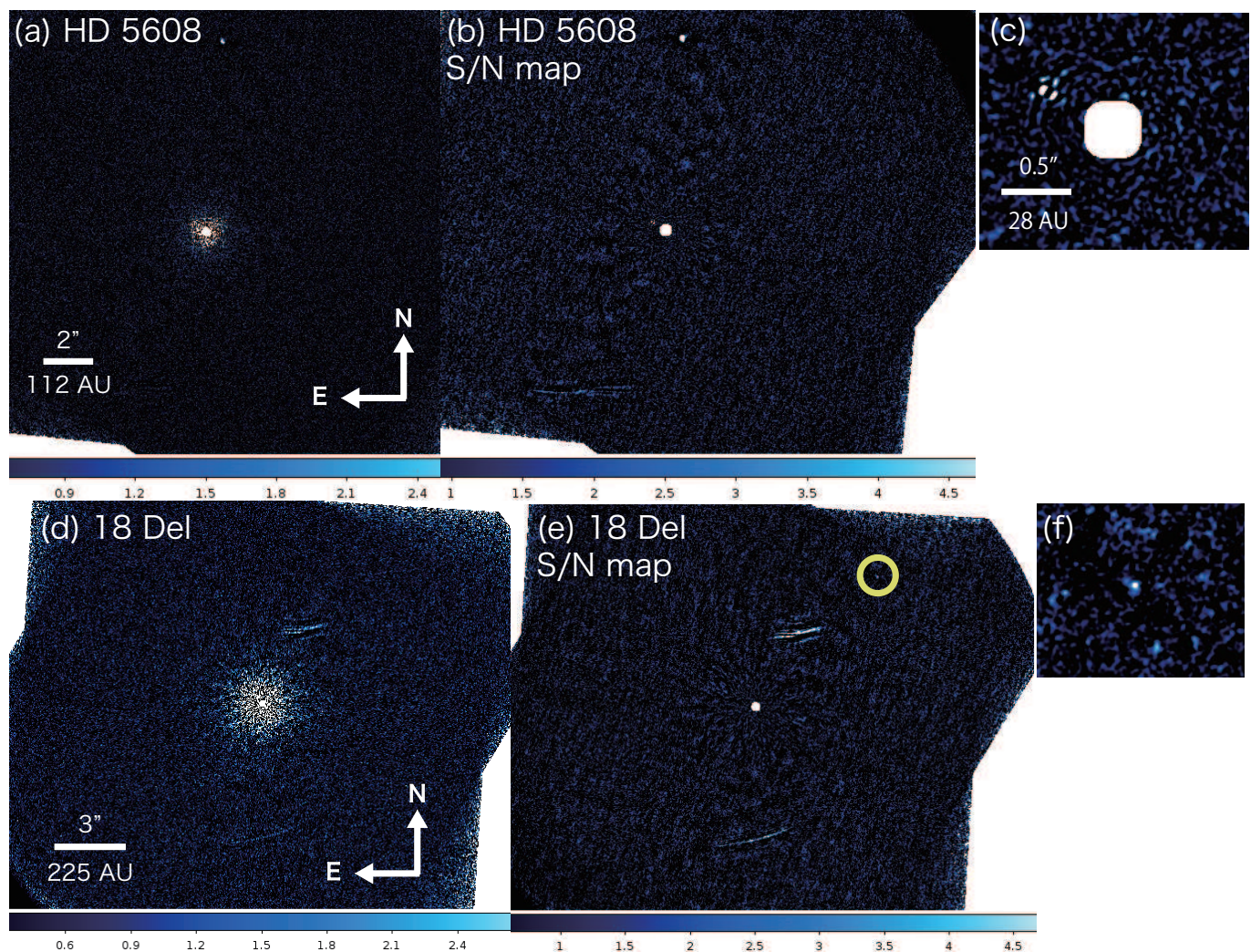}
\caption{ { Images of faint companion and background stars detected by HiCIAO observations for HD 5608 and 18 Del.  S/N map for each target is also shown to help to see a faint companion candidate, which is difficult to see on the panel (a) or (d).  (a) Final image of HD 5608 in the H-band taken on 2011 December 31. 
(b) S/N map of HD 5608 at H-band showing faint companion candidates.  Two companion candidates can be distinguished from the noise. A close faint companion candidate can be seen 0.6 from the central star. A distant companion candidate is detected 7.4 from the central star. (c) Closed-up S/N map showing the inner candidate of HD 5608. (d) Final image of 18 Del in the H-band taken on 2011 August 2. (e) S/N map of 18 Del at H-band showing a faint companion candidate.
}
}
\label{hiciaoresult2}
\end{figure}

We found two companion candidates around HD 5608. { The first candidate has $\Delta$H = 9.40 with a separation of 0\farcs6 and the second has $\Delta$H = 13.1 with a separation of 7\farcs4, as shown in Figure \ref{hiciaoresult2}. The time intervals of our three observations are long enough to allow for common proper motion tests (Figure \ref{cpmresult}).} We conclude that the close companion candidate HD 5608 B is co-moving and the other companion candidate at 7\farcs4 is a background star. Our astrometric and photometric results for HD 5608 B are shown in Table \ref{hr275bresult}. { Considering that HD 5608 B has an age of 2.5 Gyr \citep[][]{takeda08} }, we used the interpolation between the 2-Gyr and 3-Gyr Dusty models  to estimate the mass of HD 5608 B. The mass derived from Dusty is $0.106 \pm 0.002 
~ M_\odot$ from the weighted mean of three observational results, and that derived from NextGen model is $0.13  \pm 0.01 ~ M_\odot$. These indicate that HD 5608 B is a low mass star, { and the Dusty model is a model to reproduce the luminocity of the low-mass stars	.} Thus, we adopted the interpolation between the 2-Gyr and 3-Gyr Dusty models, and took the mass of HD 5608 B to be $0.10 \pm 0.01 ~ M_\odot$.

\subsubsection{HD 109272}
We discovered one companion candidate ($\Delta$H = 7.18) at a separation of 1\farcs2 from HD 109272 (Figure \ref{hiciaoresult1}). Follow-up observations allowed a common proper motion test for the candidate (Figure \ref{cpmresult}), which { suggested that} the companion candidate HD 109272 B is gravitationally bound to the central star. Table \ref{hr4779bresult} shows the astrometric and photometric results for HD 109272 B. Its mass was calculated using the 1-Gyr Dusty model {{based on an age of HD 109272 B of 1.4 Gyr \citep{takeda08}}}. { By averaging two mass estimates derived from two observations,} we find that HD 109272 B has a mass of $0.28 \pm 0.06 ~ M_\odot$.

\subsection{Confirmed background star}\label{backgroundcase}
\subsubsection{18 Del}
\begin{figure}
\plottwo{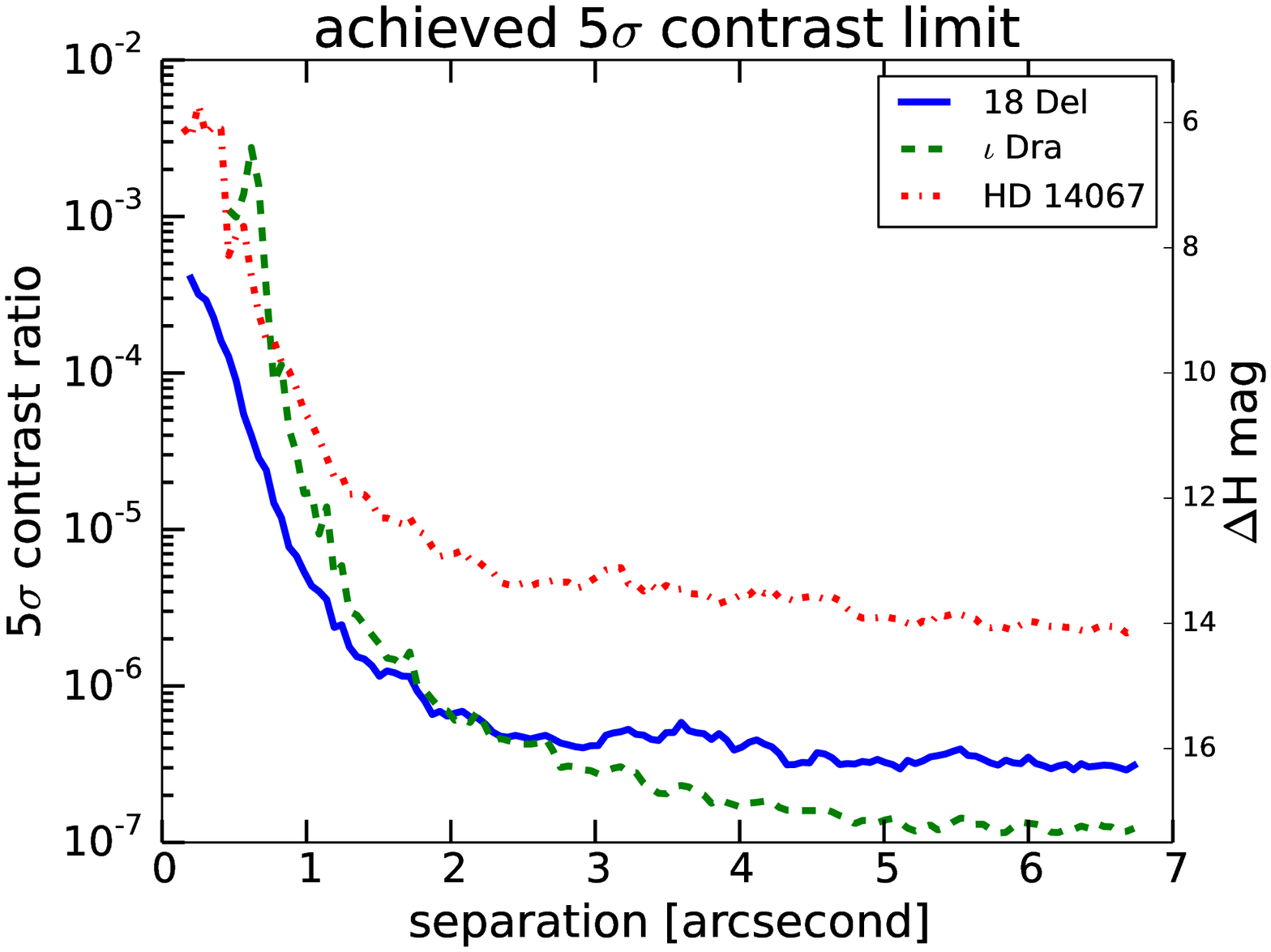}{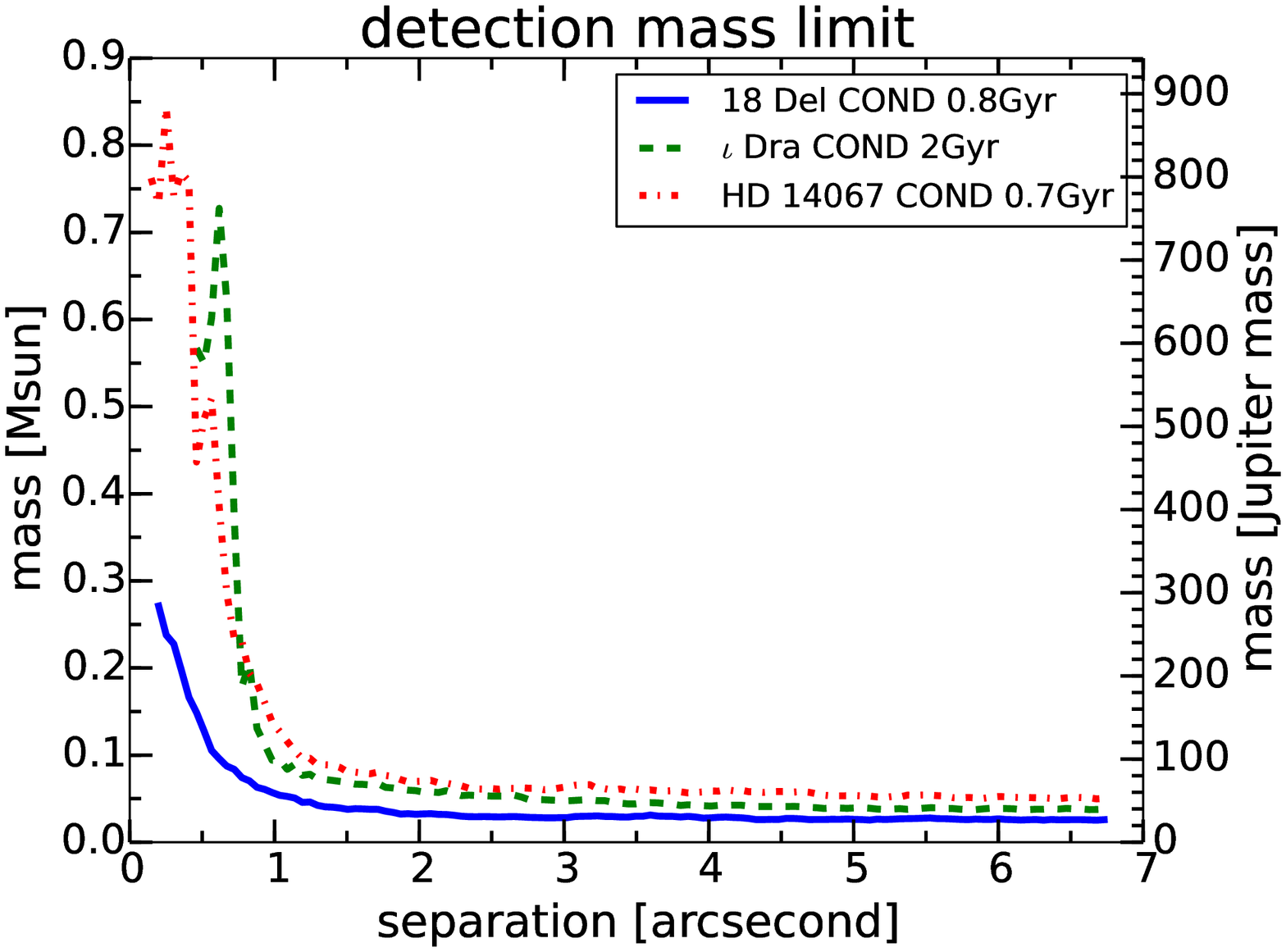}
\caption{Left: Achieved 5$\sigma$ contrast ratio on 2011 Aug.\ 2 for 18 Del in the H-band, 2012 May 14 for $\iota$ Dra in the H-band, and 1012 Nov.\ 5 for HD 14067 in the H-band. Right: Detectable mass limits for 18 Del, $\iota$ Dra, and HD 14067.}
\label{contrastandmasslimit}
\end{figure}

We found a faint companion candidate of $\Delta$H = 16.9 with S/N $\sim$ 5 at 7\farcs5 from 18 Del A (Figure \ref{hiciaoresult2}). In the July 2012 observation, we { were not able to detect the candidate because the exposure time was not enough to detect the candidate.} We detected it again on 2014 June 10th and we { carried out} a common proper motion test. The result indicates that the companion candidate traces { the track expected for a background star} (Figure \ref{cpmresult}). The achieved contrast ratio is shown in Figure \ref{contrastandmasslimit}. The detectable mass limits {{derived from}} the COND 0.8-Gyr model \citep{baraffe03} {{for an age of 18 Del of 0.79 Gyr \citep{takeda08} }} is displayed in the right panel of Figure \ref{contrastandmasslimit}. We exclude a $\sim 0.13 ~ M_\odot $ object at 0\farcs5, a $\sim 0.05 ~ M_\odot$ object at 1\farcs0,  and a $\sim0.03 ~ M_\odot \approx 31 ~ M_\mathrm{Jup}$ object beyond 2\farcs0 from the central star.

\subsection{No companion candidate detections}\label{nodetectioncase}
\subsubsection{$\iota$ Dra}
\begin{figure}
\plotone{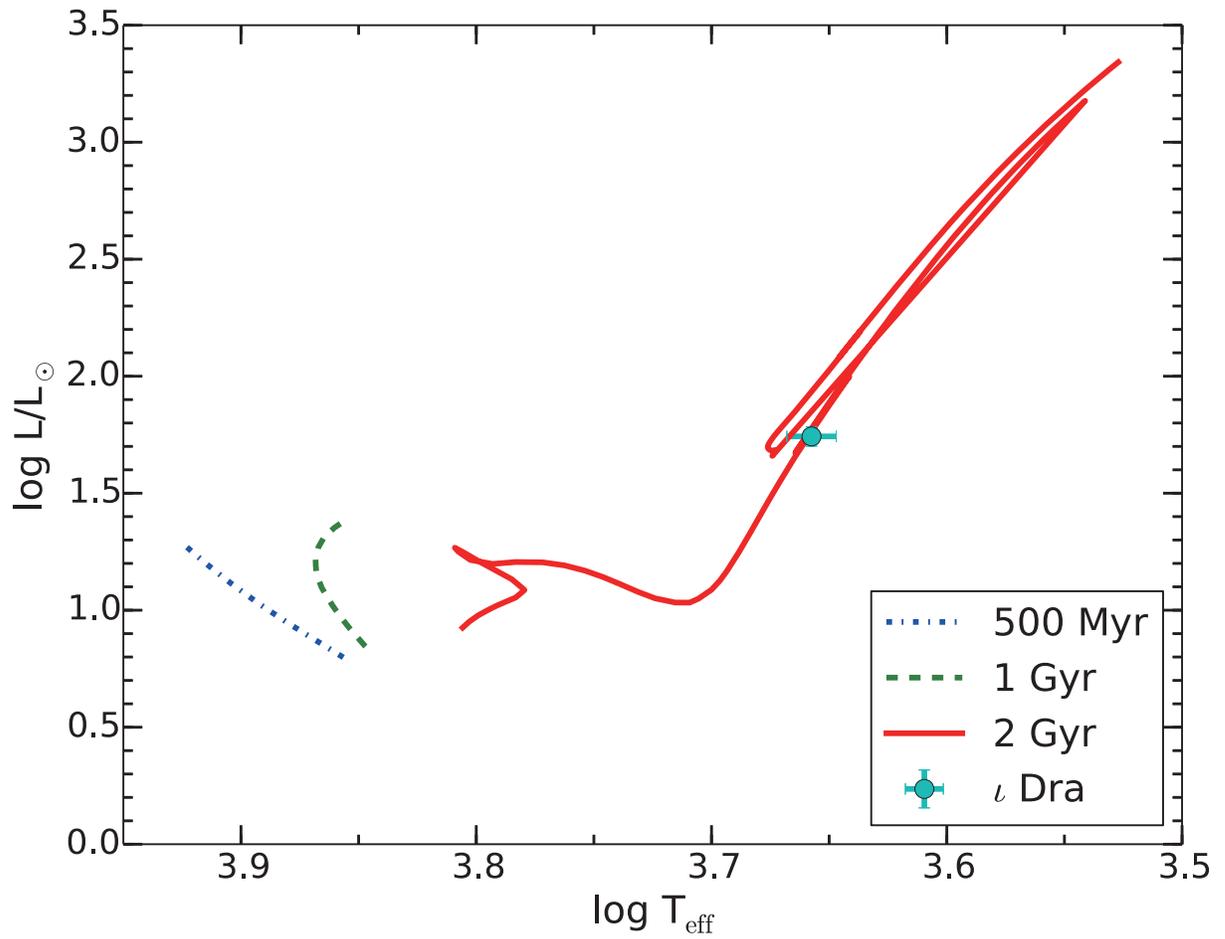}
\centering
\caption{PARSEC \citep{bressan12} isochrone plot, effective temperature $T_\mathrm{eff}$ vs. luminosity. 
The green dot with error bars is the measured value for $\iota$ Dra, which clearly agrees with the 2-Gyr PARSEC isochrone model. }
\label{ageestimate}
\end{figure}

We were not able to detect any objects on 2012 May 14 in the H-band for $\iota$ Dra beyond 0\farcs6. 
Figure \ref{ageestimate} plots the PARSEC isochrone \citep{bressan12} model within the range of the {{uncertainty}} of $\iota$ Dra's mass derived by \citet{baines11}. The position of $\iota$ Dra in the figure agrees with an age of 2 Gyr. Hence we use the 2-Gyr COND model \citep{baraffe98} to evaluate the detectable mass limits. { Although the age estimation for iot Dra may not be accurate, we note that there is not a big difference in the results even if the difference in the adopted age is $\pm$ 1 Gyr. }
 The {{excluded}} object mass range is shown in Figure \ref{contrastandmasslimit}. We exclude a $\sim 0.09 ~ M_\odot$ object at 1\farcs0 and a $\sim 0.05 ~ M_\odot \approx 52 ~ M_\mathrm{Jup}$ object over 2\arcsec.

\subsubsection{HD 14067}
We were not able to find any companion candidates around HD 14067. Figure \ref{contrastandmasslimit} shows the 5$\sigma$ detectable mass limit, converted using the COND 0.7-Gyr model {{for the age of HD 14067 \citep{wang14}}}. No objects with $\sim 0.12 ~ M_\odot $ at 1\farcs0 and $\sim 0.06 ~ M_\odot$ beyond 2\farcs0 are apparent in the observation.

\section{Discussion}
\subsection{RV trend generators}\label{causalobjects}
{{Combining the RV trend for the primary star with the projected separation of the detected companions, we can calculate the minimum dynamical mass that would be required to produce the RV trend with the following equation \citep{torres99,liu02}: }}
\begin{equation}
M_\mathrm{dyn} =5.34 \times 10^{-6} \, M_\odot \left( \frac{d}{\mathrm{pc} } \frac{\rho}{\mathrm{arcsec}} \right)^2 \times \left| \frac{\dot{v}}{\mathrm{m/s/yr}} \right| F\left(i,e,\omega,\phi \right) \label{torres99eq}
\end{equation}
where $d$ is the distance to the target, $\rho$ is the observed angular separation of the companion \citep[see also][]{knutson14}, and $F$ is a function that depends on the orbital parameters (inclination $i$, eccentricity $e$, longitude of periastron $\omega$, and orbital phase $\phi$) of the companion. \citet{torres99} determined that the minimum value of $F\left(i,e,\omega,\phi \right)$ is $3\sqrt{3}/2$. We use this equation to {{calculate the}} minimum mass limited by the RV trend. If the mass estimated {{from}} photometry exceeds the dynamical minimum mass derived from the RV trend, then we can conclude that a detected companion is responsible for the observed RV trend, and if not, the companion { is not responsible for} the observed RV trend.  { Additionally, we calculate a physical (unprojected) separation of the detected companion from the central star, consulting \citet{Howard10} who derived a true separation of a stellar companion around HD 126614 by combining the companion's estimated mass with the central star's RV trend.  
 There are two solutions for each of the three companions. }
As a simplification, we do not consider projection effects on the orbit, but assume that the projected separation is equal to the semi-major axis when comparing imaging and RV limits. As shown in e.g. \citet{Brandeker2006}, the statistical mean conversion factor between the semi-major axis and the projected separation is close to 1 for eccentricity distributions representative of wide binaries, which supports this approximation.

\begin{figure}
\plotone{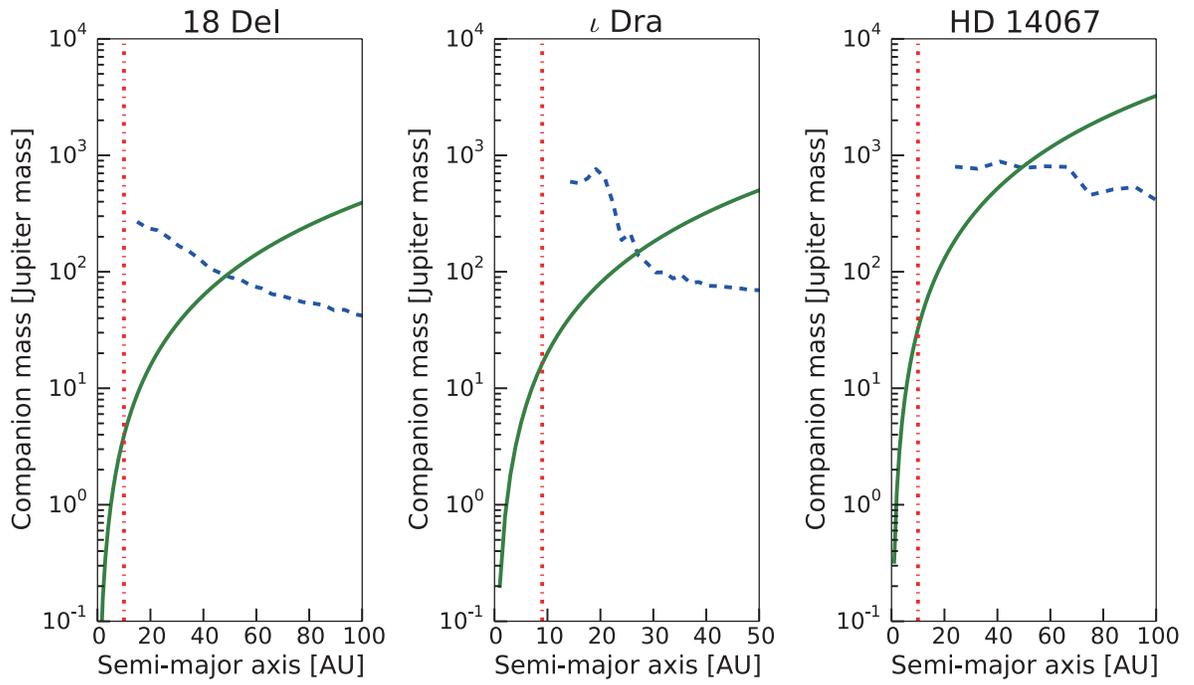}
\caption{Combined analysis from the RV trend and direct imaging data for  18 Del, $\iota$ Dra, and HD 14067. The green line is the dynamical minimum mass derived from the observed RV trend, the blue dash line is the detectable mass limit from HiCIAO observation, and the red dot-and-dash line is the limit from the observational period of the RV observations.}
\label{rvlimit}
\end{figure}

In the cases where no companion could be found, upper and lower limits for the companion as function of semi-major axis were calculated (see Figure \ref{rvlimit}). The lower mass limit from the RV trend is calculated from Equation \ref{torres99eq}, and the upper mass limit is set by the detectable mass limit of the direct-imaging observation. The object that could cause the RV trend is then constrained between these two limits.

Furthermore, the lack of curvature in a linear trend can be used to exclude the existence of inner companions. We assume that the time span of the observations must correspond to at most half an orbital period, or significant curvature would necessarily be seen. This sets a lower limit on the period, and thus an inner limit on the semi-major axis of the companion.

{\it{ $\gamma$ Hya}} -- With an angular separation of 1\farcs623 (2012 May 13), the dynamical minimum mass of $\gamma$ Hya B is 0.25 $M_\odot$. The mass estimated from photometry, $0.61^{+0.12}_{-0.14} ~ M_\odot$, exceeds the dynamical minimum mass. Therefore, we conclude that $\gamma$ Hya B is responsible for the observed RV trend. { The physical separation is 67.5 $\pm$ 0.6 AU or 159 $\pm$ 7 AU.}

{ \it{HD 5608}} -- The dynamical minimum mass of HD 5608 B is 0.095 $M_\odot$ (2011 Dec.\ 31). Our photometric estimated mass is 0.10 $\pm 0.01 \, M_\odot$, which is consistent with the dynamical minimum mass derived from the RV trend. We confirm that the companion is the RV trend generator.
{The calculated physical separation is 40 $\pm$ 1 AU or 47 $\pm$ 3 AU.}

{ \it{HD 109272}} -- In the HD 109272 system, the dynamical minimum mass limit calculated with the angular separation on 2012 Apr.\ 11 is 0.12 $M_\odot$. The estimated mass from the photometry of HD 109272 B is 0.28$ \pm 0.06 \, M_\odot$. Therefore, we conclude that HD 109272 B is the {{RV trend generator}} for the observed RV trend in HD 109272 A. 
{The true physical separation of HD 109272 B is 59.3 $\pm$ 0.9 AU or 140 $\pm$ 6 AU.}

{ \it{18 Del}} -- \citet{mugrauer14} reported that 18 Del A has a distant companion 18 Del B outside the field of view of HiCIAO. The projected separation of 18 Del B is 2199 AU and its mass is 0.19 $M_\odot$. The dynamical minimum mass at 2199 AU is 181 $M_\odot$. Therefore, 18 Del B cannot be the source of the observed RV trend. The upper and lower limits for the {{RVTG}} are shown in Figure \ref{rvlimit}. In addition, the long-term linear RV trend would excludes the existence of inner objects. The semi-major axis range of the RVTG is {{$a \sim$}} 10--50 AU. A stellar companion at wide separation is ruled out, though a low-mass stellar companion at the inner region is possible. The minimum mass at {{$a=$}} 10 AU is {{$m_p \approx$}} 4 $M_\mathrm{Jup}$, so the RVTG is either a high-mass planet, a brown dwarf, or a low-mass stellar companion.

{ \it{$\iota$ Dra}} -- The absence of the detection of any companions around $\iota$ Dra is consistent with the result observed by \citet{kane14} at 692 nm and 880 nm. 
An analysis combining the RV trend and the HiCIAO result is shown in Figure \ref{rvlimit}.  Considering the linear RV trend observed over a decade \citep{kane10}, the possible innermost object is {{$m_p \approx$}} 16 $M_\mathrm{Jup}$ at{{$a =$}} 9 AU. On figure \ref{rvlimit}, the intersection of the observation detectable line with the dynamical minimum line derived from the RV trend is {{$a \approx$}} 31 AU. Hence, the semi-major axis range of the RVTG is {{$a \sim$}} 9--31 AU. The mass range implies the RVTG is a brown dwarf or a stellar companion at small separation.

{ \it{HD 14067}} -- From the observation result, we can determine limits for identifying the RVTG for HD 14067 (Figure \ref{rvlimit}). The linear RV trend over five years excludes objects in the inner region. The possible objects' orbital period is 10 years at least, which means that the innermost possible {{RVTG}} is at {{$a=$}} 10 AU in the HD 14067 system. The minimum dynamical mass at $a=$ 10 AU in the system is $m_p \approx$ 32 $M_\mathrm{Jup}$. On figure \ref{rvlimit}, the outermost possible RVTG is at $a \approx$ 49 AU, where the detectable limit line crosses the dynamical minimum line. The dynamical minimum mass at $a \approx$ 49 AU is $M_p \approx$ 0.74 $M_\odot$. We can rule out a wide-orbit stellar companion, though a stellar companion at $a \sim$ 10--49 AU is still possible. The RVTG for HD 14067 is a brown dwarf or a stellar companion at $a \sim$ 10--49 AU.

Our high-contrast observations would exclude planetary {{RVTGs for five out of the six targets}}. The exception, 18 Del, can be either a planet or a low-mass star. To distinguish the nature of the RVTG of 18 Del, further RV monitoring and higher contrast imaging for the inner region using extreme AO (e.g.,  SCExAO; \citealt[][]{Martinache09}) would be important.

Two systems, $\iota$ Dra and HD 14067, for which we cannot identify RVTGs in the high-contrast imaging, are important. \citet{vigan12} reported that the frequency of brown dwarfs about intermediate-mass stars is as low as 2.8$^{+6.0}_{-0.9}$\% at {{a range of separation}} of 5--320 AU. {{On the one hand,}} \citet{duchene13} reported that the {{orbital period distribution for intermediate-mass multiple systems}} has two peaks at $P\approx$ 10 days and $a \approx$ 350 AU. In either case, the two systems offer unique samples of brown dwarfs or low-mass stars around intermediate-mass stars. Further high-contrast imaging observations with deeper and inner sensitivity would be important not only to clarify the frequency of brown dwarfs and low-mass stars around intermediate-mass stars, but also to study the planet migration of the inner eccentric planets in those systems.

\subsection{Mechanism influencing the orbit of inner eccentric planets}\label{Kozaisection}

Several studies have revealed that the formation mechanism of eccentric planets cannot be explained by core accretion theory and Type I/II migration. The Kozai mechanism, which is a perturbation mechanism from a distant stellar companion \citep[e.g.][]{wu03}, planet--planet scattering \citep[e.g.][]{nagasawa08}, and secular chaotic excursions \citep[e.g.][]{wu10} are promising approaches to describe eccentric planets. 

Four targets have already known inner planets (Table \ref{innerplanets}). 
We consider the Kozai mechanism to explain the eccentric planets, namely a perturbation due to an outer stellar companion periodically oscillates the eccentricity and inclination of an inner planet. The oscillation timescale of the Kozai mechanism is calculated by
\begin{equation}
\label{eqKozai}
P_\mathrm{Kozai} \sim \frac{M_A}{m_B} \frac{P_{B}^{2}}{P_{b,0}} \left( 1- e_B^2 \right)^{3/2} 
\end{equation}
where $M_A$ is the primary star's mass, $m_B$ is the stellar companion's mass, $P_B$ is the period of the companion, $P_{b,0}$ is the initial period of the planet, and $e_B$ is the eccentricity of the companion \citep{holman97}.

We calculated the timescale of the Kozai mechanism for the significantly high eccentricity planets $\iota$ Dra b, HD 5608 b, and HD 14067 b. For $\iota$ Dra b, we assume that its companion has a mass of 0.18 $M_{\rm{\odot}}$ and in a circular orbit at 31 AU, {{which}} is the maximum mass and separation.
We then assume that the initial period of the planet is equal to that of a circular orbit of the observed semi-major axis. The timescale is $P_\mathrm{Kozai} \sim$ { 107 kyr}, which is much less than the Gyr order of $\iota$ Dra's age. { The timescale for HD 5608 b, assuming the circular orbit of HD 5608 B, is $P_\mathrm{Kozai} \sim$ 450 kyr or 277 kyr.} The timescale for HD 14067 b is $P_\mathrm{Kozai} \sim$ { 30 kyr }with the 0.74 $M_\odot$ object at 49 AU. These timescales are also sufficiently shorter than the system's age. {{It follows from equation (\ref{eqKozai}) that if the planets have migrated inward from an initially wider separation, then the initial Kozai timescales would be even shorter.}} Therefore, we conclude that the Kozai mechanism could be a plausible explanation for the eccentricity of the planets $\iota$ Dra b, HD 5608 b, and HD 14067 b. We note that an alternative mechanism for producing high eccentricities is planet-planet scattering (e.g. \citealt[][]{nagasawa08}). This possibility can be tested by continuing RV and direct imaging observations, in order to search for additional planets in the systems, as well as providing improved constraints for the orbits of the imaged companions.

\section{Conclusion}\label{conclusion}
We present direct-imaging results of intermediate-mass stars with long-term RV trends that indicate the existence of an outer object. We used the HiCIAO/Subaru Telescope to identify the objects responsible for the observed RV trends. Our observations revealed that the three evolved intermediate-mass stars $\gamma$ Hya, HD 5608, and HD 109272 possess the stellar companions $\gamma$ Hya B, HD 5608 B, and HD 109272 B, respectively. We also ruled out the presence of { stellar companions and} brown dwarfs for separations from 1$\arcsec$ to 7$\arcsec$ for $\iota$ Dra, 18 Del, and HD 14067.

{ {We have constrained the nature of the RVTGs around each of the six targets.}} The detected companions $\gamma$ Hya B, HD 5608 B, and HD 109272 B exceed the minimum dynamical mass derived from the combination of RV and direct imaging observations. We confirm that these companions are responsible for the observed RV {{trends}}. We also calculated the upper and lower limits {{of}} the mass and the {{semi-major axis}} for the {{RVTGs of}}  $\iota$ Dra, 18 Del, and HD 14067. These RVTGs are promising candidates for hosting brown dwarfs or possibly low-mass stellar companions.

The existence of the companions around eccentric planet systems suggests that the Kozai mechanism is a plausible explanation for the eccentricity. For the three eccentric planet systems, $\iota$ Dra b, HD 5608 b, and HD 14067 b, the Kozai oscillation timescales are significantly shorter than their age, and thus the Kozai mechanism is a plausible explanation for the eccentricity of the planets.

\acknowledgments

The authors thank David Lafreni\`ere for generously providing the source code for the LOCI algorithm. This paper is based on data collected at the Subaru Telescope and the 1.88-m telescope at OAO, operated by the National Astronomical Observatory of Japan. We thank the observatory staff for their special support during the HiCIAO, AO188, and HIDES observations.

The data analysis was carried out using a common use data analysis computer system at the Astronomy Data Center of the National Astronomical Observatory of Japan. This research made use of the SIMBAD database, operated at CDS, Strasbourg, France. 
Our analysis is also based on observations made with the NASA/ESA Hubble Space Telescope, and obtained from the Hubble Legacy Archive, which is a collaboration between the Space Telescope Science Institute, the Space Telescope European Coordinating Facility (ST-ECF/ESA) and the Canadian Astronomy Data Centre (CADC/NRC/CSA). N.N. acknowledges supports by the NAOJ Fellowship, Inoue Science Research Award, and a Grant-in-Aid for Scientific Research (A) (JSPS KAKENHI Grant Number 25247026). J.C.C. acknowledges support by the U.S. National Science Foundation under Award No. 1009203. This work was partially supported by a Grant-in-Aid for JSPS Fellows (Grant Number 25-8826). 

The authors wish to recognize and acknowledge the very significant cultural role and reverence that the summit of Mauna Kea has always had within the indigenous Hawaiian community. We are most fortunate to have the opportunity to conduct observations on this mountain.

\begin{deluxetable}{ccccccc}
\tabletypesize{\scriptsize}
\tablewidth{0pt}
\tablecaption{Stellar properties of targets\label{targetproperty} }
\tablehead{
\colhead{Property}&\colhead{$\gamma$ Hya}&\colhead{$\iota$ Dra}&\colhead{18 Del}&\colhead{HD 5608}&\colhead{HD 14067}&\colhead{HD 109272}} 
\startdata
Other name&HD 115659&HD 137759&HD 199665&HR 275&HR 665&HR 4779\\
R.A. (J2000) \tablenotemark{a} & 13:18:55.297 &15:24:55.775&20:58:25.934&00:58:14.219&02:17:10.440&12:33:34.258\\
Dec. (J2000) \tablenotemark{a} & -23:10:17.45&+58:57:57.83&+10:50:21.43&+33:57:03.18&+23:46:04.18&-12:49:48.73\\
J (mag) \tablenotemark{b} & 1.519 $\pm$ 0.278 & 1.293 $\pm$ 0.220 & \nodata & \nodata & 4.718 $\pm$ 0.037 &4.151 $\pm$ 0.280\\
H (mag) \tablenotemark{b} & 1.065 $\pm$ 0.266 &0.724 $\pm$ 0.146& 3.44 $\pm$ 0.08 & 3.89 $\pm$ 0.05& 4.448 $\pm$ 0.220 &3.616 $\pm$ 0.226\\
K (mag) \tablenotemark{b} & 1.024 $\pm$ 0.300 &0.671 $\pm$ 0.200 &\nodata & \nodata & 4.097 $\pm$ 0.036 &3.600 $\pm$ 0.250\\
Distance (pc) \tablenotemark{c} &41.0 $\pm$ 0.2& 31.0 $\pm$ 0.1 &75 $\pm$ 1& 56 $\pm$ 1 &163 $\pm$ 13 &49.3 $\pm$ 0.8\\
$\mu_\alpha$ (mas/yr) \tablenotemark{a} &68.99 $\pm$ 0.17 & -8.36 $\pm$ 0.08 &-48.75 $\pm$ 0.33 &34.98 $\pm$ 0.40 &-32.57 $\pm$ 0.48 &-17.64 $\pm$ 0.28\\
$\mu_\delta$ (mas/yr) \tablenotemark{a} &-41.85 $\pm$ 0.09 & 17.08 $\pm$ 0.10& -34.43 $\pm$ 0.17 &-71.87 $\pm$ 0.20& -42.21 $\pm$ 0.44& 52.09 $\pm$ 0.19\\
Mass ($M_\odot$) &2.94 $^{+0.03}_{-0.06}$ \tablenotemark{d}&1.82 $\pm$ 0.23 \tablenotemark{e} & 2.25 $^{+0.05}_{-0.06}$ \tablenotemark{d} & 1.55 $\pm$ 0.11 \tablenotemark{d} & 2.4 $\pm$0.2 \tablenotemark{f} & 1.79 $\pm$ 0.11 \tablenotemark{d} \\
Sp.\ type & G8III & K2III & G6III & K0IV & G9III & G8III/IV \\
$[$Fe/H$]$  & -0.04 $\pm$ 0.04 \tablenotemark{d} & 0.07 $\pm$ 0.08 \tablenotemark{g} &-0.05 $\pm$ 0.04 \tablenotemark{d} & 0.06 $\pm$ 0.05 \tablenotemark{d} & -0.10 $\pm$ 0.08 \tablenotemark{f} & -0.26 $\pm$ 0.02 \tablenotemark{d}\\
$T_\mathrm{eff}$ (K) &5019 $\pm$ 20 \tablenotemark{d} & 4545 $\pm$ 110 \tablenotemark{e} & 4985 $\pm$ 18 \tablenotemark{d} & 4854 $\pm$ 25 \tablenotemark{d} & 4815 $\pm$ 100 \tablenotemark{f} & 5104 $\pm$ 10 \tablenotemark{d} \\
Age (Gyr) & 0.37 {{$^{+0.03}_{-0.01}$}} \tablenotemark{d} & \nodata & 0.79 {{$\pm$ 0.05}} \tablenotemark{d} & 2.5 {{$^{+1.4}_{-1.0}$}} \tablenotemark{d} &0.69 $\pm$ 0.20 \tablenotemark{f} & {{ 1.4 $^{+0.3}_{-0.1}$}} \tablenotemark{d}\\
\enddata
\tablenotetext{a}{Refined data reduction of {\it Hipparcos} \citep{van07}}
\tablenotetext{b}{Calibrated by stdstar in this work}
\tablenotetext{c}{The parallax-based distance from {\it Hipparcos} uses \citet{van07}}
\tablenotetext{d}{\citet{takeda08}}
\tablenotetext{e}{\citet{baines11}}
\tablenotetext{f}{\citet{wang14}}
\tablenotetext{g}{\citet{dasilva11}}
\end{deluxetable}

\begin{deluxetable}{ccccccc}
\tablewidth{0pt}
\rotate
\tabletypesize{\scriptsize}
\tablecaption{Summary of known planets \label{innerplanets}}
\tablehead{\colhead{Name} &\colhead{{{Minimum planetary}} mass ($M_\mathrm{Jup}$ )} &\colhead{Period (days)} &\colhead{Semi-major axis (AU) } &\colhead{Eccentricity} &\colhead{Periastron separation (AU)} &\colhead{Stellar companion?} }
\startdata
$\iota$ Dra b & 12 $\pm$ 1.1 \tablenotemark{a} & 510.72 $\pm$ 0.07 \tablenotemark{b} & 1.27 \tablenotemark{b} & 0.713 $\pm$ 0.008 \tablenotemark{b} & 0.36 & No \tablenotemark{c,d}\\
18 Del b & 10.3 \tablenotemark{e} & 993.3 $\pm$ 3.2 \tablenotemark{e} & 2.6 \tablenotemark{e}& 0.08 $\pm$ 0.01 \tablenotemark{e} & 2.4 & Yes \tablenotemark{f}\\
HD 5608 b & 1.4 \tablenotemark{g} & 792.6 $\pm$ 7.7 \tablenotemark{g} & 1.9 \tablenotemark{g} & 0.190 $\pm$ 0.061 \tablenotemark{g} &1.5& Yes \tablenotemark{d}\\
HD 14067 b & 7.8 $\pm$ 0.7 \tablenotemark{h} & 1455 $^{+13}_{-12}$ \tablenotemark{h} & 3.4 $\pm$ 0.1 \tablenotemark{h} & 0.533 $^{+0.043}_{-0.047}$ \tablenotemark{h} &1.6& No \tablenotemark{d}
\enddata
\tablenotetext{a}{\citet{baines11}} 
\tablenotetext{b}{\citet{kane10}}
\tablenotetext{c}{\citet{kane14}}
\tablenotetext{d}{This work}
\tablenotetext{e}{\citet{sato08}}
\tablenotetext{f}{\citet{mugrauer14}}
\tablenotetext{g}{\citet{sato12}}
\tablenotetext{h}{\citet{wang14}}
\end{deluxetable}

\begin{deluxetable}{lrr}
\tabletypesize{\scriptsize}
\tablecolumns{3}
\tablewidth{0pt}
\tablecaption{Updated radial velocities for 18 Del \label{tbl:18Del-RV} }
\tablehead{
\colhead{JD-2450000} & \colhead{Velocity (\ms)} & \colhead{Uncertainty
(\ms)}}
\startdata
2489.14222 & 11.13 & 4.30\\
2507.12660 & $-$30.85 & 4.42\\
2541.12604 & $-$54.77 & 4.08\\
2857.13560 & $-$40.08 & 5.34\\
2896.04035 & $-$16.81 & 4.88\\
2927.05176 & $-$6.01 & 4.10\\
2974.90124 & 22.45 & 4.09\\
2994.89777 & 49.79 & 4.78\\
3005.89596 & 46.26 & 7.87\\
3008.88731 & 46.14 & 5.59\\
3077.34500 & 98.80 & 6.63\\
3100.29135 & 94.18 & 5.30\\
3131.31381 & 100.95 & 5.69\\
3201.11623 & 155.35 & 4.39\\
3246.10685 & 114.78 & 4.12\\
3249.10725 & 110.37 & 3.93\\
3284.92436 & 101.52 & 5.12\\
3289.95478 & 109.83 & 4.31\\
3305.92236 & 84.59 & 3.59\\
3310.94725 & 77.87 & 3.65\\
3331.91858 & 77.27 & 3.79\\
3334.87618 & 80.06 & 3.82\\
3340.00938 & 64.08 & 5.15\\
3362.87668 & 77.89 & 3.96\\
3364.90638 & 86.57 & 4.58\\
3428.37046 & 41.14 & 5.49\\
3448.34315 & $-$5.99 & 5.47\\
3470.31682 & $-$31.43 & 11.74\\
3474.31881 & $-$38.64 & 5.88\\
3495.26235 & $-$51.75 & 6.41\\
3520.29312 & $-$40.30 & 4.72\\
3525.28825 & $-$43.20 & 11.55\\
3527.29963 & $-$46.68 & 8.18\\
3576.98990 & $-$104.36 & 8.66\\
3579.13224 & $-$110.44 & 6.20\\
3600.04881 & $-$111.32 & 9.58\\
3635.09840 & $-$99.09 & 4.66\\
3655.94667 & $-$121.89 & 3.65\\
3692.90100 & $-$124.49 & 3.79\\
3719.92105 & $-$124.78 & 4.05\\
3726.87981 & $-$130.32 & 4.52\\
3740.88187 & $-$118.68 & 5.81\\
3815.34344 & $-$87.89 & 5.75\\
3833.33315 & $-$80.06 & 5.88\\
3853.29087 & $-$61.78 & 6.74\\
3890.21908 & $-$39.35 & 8.24\\
3938.27152 & 2.55 & 4.23\\
3962.21118 & 17.32 & 4.63\\
4018.04395 & 59.08 & 3.89\\
4048.99644 & 63.22 & 4.51\\
4088.89931 & 93.57 & 3.92\\
4195.31850 & 98.01 & 4.63\\
4216.31524 & 104.88 & 5.73\\
4254.23118 & 118.57 & 4.91\\
4261.26613 & 109.80 & 4.31\\
4305.15291 & 83.07 & 4.24\\
4338.05996 & 45.84 & 4.23\\
4378.14307 & 22.67 & 6.86\\
4415.97059 & 20.46 & 4.46\\
4460.92172 & $-$51.46 & 4.12\\
4558.32454 & $-$95.51 & 5.34\\
4587.31328 & $-$98.00 & 10.00\\
4588.29756 & $-$107.60 & 5.06\\
4624.27152 & $-$135.11 & 4.90\\
4672.10848 & $-$140.78 & 3.88\\
4703.11346 & $-$117.89 & 10.70\\
4704.04561 & $-$115.06 & 4.12\\
4756.08274 & $-$94.94 & 4.24\\
4756.97318 & $-$88.86 & 3.75\\
4800.91327 & $-$96.02 & 4.06\\
4817.95988 & $-$87.37 & 5.64\\
4818.93093 & $-$83.47 & 4.67\\
4983.24038 & 36.22 & 4.44\\
5036.17085 & 125.72 & 6.51\\
5107.94026 & 98.37 & 5.12\\
5137.05576 & 99.34 & 4.42\\
5165.89151 & 126.77 & 3.62\\
5350.30105 & 47.36 & 3.99\\
5470.95791 & $-$74.04 & 4.13\\
5787.12587 & $-$104.19 & 4.77\\
6142.20352 & 107.85 & 4.31\\
6162.07563 & 101.40 & 4.15\\
\label{tbl:18Del-RV}
\enddata
\end{deluxetable}

\begin{deluxetable}{ccc}
\tablewidth{0pt}
\tablecaption{Radial velocities of $\gamma$ Hya \label{rvvaluegamhya} }
\tablehead{
\colhead{JD-2450000}&\colhead{RV (m/s)} &\colhead{Uncertainty (m/s)} }
\startdata
2312.29897 & $-$16.90& 3.73\\
2340.27905 &$-$23.01& 5.83 \\
2655.33897& $-$2.80&4.79\\
2680.31050 &$-$13.40& 2.79\\
2736.13727 &$-$6.16& 5.32\\
2783.04181& $-$21.85& 6.33\\
3053.24088 &$-$23.57 &4.28\\
3408.29633& $-$12.93& 2.75\\
3812.23474 &1.61 &7.14 \\
3812.26327& $-$1.59& 4.99\\
4110.36057& $-$5.55& 3.36\\
4492.39610& 8.25& 4.79\\
4525.24386& 4.18& 4.81\\
4863.38961& 2.64& 3.81\\
5204.39491& 11.04& 2.90\\
5347.97797& 16.89& 4.98\\
5583.37648& 29.79& 3.17\\
5614.30154& 3.61& 4.12\\
5661.12157& 25.34& 4.59\\
5923.40131& 18.42& 4.85\\
\enddata
\end{deluxetable}

\begin{deluxetable}{ccc}
\tablewidth{0pt}
\tablecaption{Radial velocities of HD 109272 \label{rvvaluehd109272} }
\tablehead{
\colhead{JD-2450000}&\colhead{RV (m/s)} &\colhead{Uncertainty (m/s)} }
\startdata
1963.20405& $-$11.23& 5.91\\
1965.23360& $-$10.55& 6.02 \\
1966.14126& $-$4.85& 5.57\\
1966.16086& $-$10.29& 5.89\\
2016.16054& $-$9.43& 6.69\\
2036.08305& $-$12.20& 5.66\\
2043.06174& $-$18.89& 4.92\\
2272.32149& $-$0.35& 4.55\\
2337.24699& 4.10& 5.95\\
2424.02042& $-$6.80& 7.09\\
2653.28434& 0.07& 5.06\\
2707.27045& $-$4.56& 11.29\\
2710.20140& $-$12.54& 5.40\\
3113.12341& 2.77& 6.29\\
3367.24880& $-$13.43& 7.43\\
3812.18616& $-$5.59& 5.13\\
4093.38219& $-$0.36& 4.38\\
4495.29519& 12.62& 5.38\\
4884.20448& 1.40& 4.35\\
5234.24683& 16.30& 5.19\\
5350.01752& 12.53& 4.95\\
5556.32922& 15.21& 5.02\\
5626.20043& 10.42& 4.46\\
5663.12343& 28.28& 5.04\\
5977.23234& 16.95& 4.75\\
\enddata
\end{deluxetable}

\begin{deluxetable}{lr}
\tablecaption{Updated orbital parameters for 18 Del\label{tbl:18Del-orbit}}
\tablewidth{0pt}
\tablehead{
\colhead{Parameter} & \colhead{18 Del b}}
\startdata
$P$ (days)                            &  982.2 $\pm$ 3.4 \\
$K_1$ (\ms)                           &  121.7 $\pm$ 2.2  \\
$e$                                   &  0.016$^{+0.017}_{-0.011}$ \\
$\omega$ (deg)                        &  $-$210$^{+84}_{-73}$   \\
$T_p$    (JD$-$2450000)             & $-$310$^{+230}_{-200}$   \\
$a_1\sin i$ (10$^{-3}$ AU)            & 10.98 $\pm$ 0.20 \\
{\large{  $f_1(m)$ }} (10$^{-7}\,\Msun$)         & 1.83$^{+0.10}_{-0.097}$  \\
$m_2\sin i$ ($\Mjup$)                    & 10.2    \\
$a$ (AU)                                           & 2.5   \\
jitter (\ms)                                           & 12.9$^{+1.3}_{-1.2}$   \\
$\dot{\gamma}$  (m s$^{-1}$ yr$^{-1}$)    & $-$2.8 $\pm$ 0.7 \\
$N_{\rm obs}$                                          & 82\\
RMS  (\ms)                                       & 13.4\\
\enddata
\end{deluxetable}


\begin{deluxetable}{ccc}
\tablewidth{0pt}
\tablecaption{RV trends of our targets \label{rvtrend_table}}
\tablehead{\colhead{Name} &\colhead{$\dot{\gamma}$ (m/s/yr) } & \colhead{} }
\startdata
$\gamma$ Hya & 4.1 $\pm$ 0.2 & This work \\
$\iota$ Dra & $-$13.65 $\pm$ 0.75 & \citet{kane10} \\
18 Del & $-$2.8 $\pm$ 0.7  & This work \\
HD 5608 & $-$5.51 $\pm$ 0.45 & \citet{sato12} \\
HD 14067  & $-$22.4 $\pm$ 2.2 & \citet{wang14} \\
HD  109272 & 2.4 $\pm$ 0.2 & This work
\enddata
\end{deluxetable}

\begin{table}
\begin{center}
\caption{Summary of observation log}
\label{obslog}
\begin{tabular}{clccccc}
\tableline\tableline
Target & Obs. date (UT) & Band & Total ET (s) & Rotation angle & Mask (\arcsec) & Notes\\
\tableline
$\gamma$ Hya & 2012 May 13& H & 585 & 10.3&0.6\\
& 2014 Apr. 25 & J & 150 & 3.0 &\nodata &cloudy\\
& 2014 Apr. 25 & H &148.5&2.8 & \nodata&cloudy\\
& 2014 Apr. 25 & Ks&75& 2.4 &\nodata &cloudy\\
\tableline
$\iota$ Dra &2012 May 14& H & 675 & 12.0&0.6&cloudy\\
\tableline
18 Del &2011 Aug. 2 & H &585&27.7&0.4 \\
&2012 Jul. 8&H&480&20.7&0.4\\
&2014 Jun. 10&H&575&31.7&\nodata\\
\tableline
HD 5608 &2011 Dec. 31&H&570&30.8&0.4\\
&2012 Sep. 12&H&966&28.1&0.6\\
&2014 Oct. 7&H&600&31.5&\nodata &cloudy\\
\tableline
HD 14067 &2012 Nov. 5&H&1600&89.3&0.4\\
\tableline
HD 109272&2012 Apr. 11&H&450&10.8&0.4 \\
&2014 Apr. 23&J& 195& 4.8& \nodata\\
&2014 Apr. 23&H&147&3.6 &\nodata\\
&2014 Apr. 23&Ks& 165& 1.0 &\nodata\\
\tableline
\end{tabular}
\end{center}
\end{table}

\begin{deluxetable}{ccccccc}
\tablewidth{0pt}
\tablecaption{Astrometric and photometric results for $\gamma$ Hya B \label{gamhyabresult}}
\tablehead{\colhead{Name}& \colhead{Date (UT)} &\colhead{Filter} &\colhead{Sep. (\arcsec) } &\colhead{P.A. (deg)} & \colhead{$\Delta$mag}  &\colhead{Mass ($M_\odot$)}}
\startdata
$\gamma$ Hya B & 2012 May 13 & H & 1.623 $\pm$ 0.011 & 194.4 $\pm$ 0.2 & 7.24 $\pm$ 0.08 & 0.63 $\pm$ 0.06\\
& 2014 Apr. 25 & J &1.611 $\pm$ 0.004  & 195.2 $\pm$ 0.2 & 7.97 $\pm$ 0.24 &0.53 $\pm$ 0.14 \\
&2014 Apr. 25 &H & 1.611 $\pm$ 0.004 & 195.2 $\pm$ 0.2 & 7.39 $\pm$ 0.20 &0.61 $\pm$ 0.12\\
&2014 Apr. 25 &Ks&1.626 $\pm$ 0.006 & 195.3 $\pm$ 0.1 & 7.07 $\pm$ 0.30 & 0.65 $\pm$ 0.21
\enddata
\end{deluxetable}

\begin{deluxetable}{clcccccc}
\tablewidth{0pt}
\tablecaption{Astrometric and photometric results for HD 5608 B \label{hr275bresult}}
\tablehead{\colhead{Name}& \colhead{Date (UT)} &\colhead{Filter} &\colhead{Sep. (\arcsec) } &\colhead{P.A. (deg)} & \colhead{$\Delta$mag} &\colhead{Mass ($M_\odot$)}}
\startdata
HD 5608 B & 2011 Dec. 31 & H & 0.627 $\pm$ 0.009 & 58.9 $\pm$ 0.4 & 9.40 $\pm$ 0.11 & 0.11 $\pm$ 0.02\\
&2012 Sep. 12 &H & 0.627 $\pm$ 0.022 & 59.9 $\pm$ 1.0 &9.70 $\pm$ 0.10 & 0.10 $\pm$ 0.02 \\
&2014 Oct. 7 &H & 0.588 $\pm$ 0.012 & 55.7 $\pm$ 0.6 & 9.55 $\pm$ 0.20 & 0.11 $\pm$ 0.02
\enddata
\end{deluxetable}

\begin{deluxetable}{cccccccc}
\tablewidth{0pt}
\tablecaption{Astrometric and photometric results for HD 109272 B \label{hr4779bresult}}
\tablehead{\colhead{Name}& \colhead{Date (UT)} &\colhead{Filter} &\colhead{Sep. (\arcsec) } &\colhead{P.A. (deg)} & \colhead{$\Delta$mag}  &\colhead{Mass ($M_\odot$)}}
\startdata
HD 109272 B & 2012 Apr. 11 & H & 1.187 $\pm$ 0.005 & 53.0 $\pm$ 0.2 & 7.18 $\pm$ 0.14 & 0.30 $\pm$ 0.04\\
& 2014 Apr. 23 & J & 1.168 $\pm$ 0.004 & 52.0 $\pm$ 0.1 & 7.36 $\pm$ 0.33 &0.28 $\pm$ 0.06 \\
&2014 Apr. 23 &H &1.166 $\pm$ 0.004 & 52.2 $\pm$ 0.1 & 7.22 $\pm$ 0.28 & 0.30 $\pm$ 0.09\\
&2014 Apr. 23 &Ks& 1.170 $\pm$ 0.006 & 52.0 $\pm$ 0.1 & 7.40 $\pm$ 0.16 &  0.24 $\pm$ 0.04
\enddata
\end{deluxetable}

\clearpage

\begin{thebibliography}{}
\bibitem[Baines et al.(2011)]{baines11}Baines, E. K., McAlister, H. A., ten Brummelaar, T. A., et al. 2011,  ApJ, 743, 130B
\bibitem[Baraffe et al.(1998)]{baraffe98}Baraffe, I., Chabrier, G., Allard, F., \& Hauschildt, P. H. 1998, A\&A 337, 403
\bibitem[Baraffe et al.(2003)]{baraffe03}Baraffe, I., Chabrier, G., Barman, T. S., et al. 2003,  A\&A, 402, 701B
\bibitem[Brandeker et al.(2006)]{Brandeker2006}Brandeker, A., Jayawardhana, R., Khavari, P., Haisch, K. E., Jr., \& Mardones, D. 2006, ApJ, 652, 1572
\bibitem[Brandt et al.(2013)]{brandt13}Brandt, T. D., McElwain, M. W., Turner, E. L., et al. 2013, ApJ, 764, 183
\bibitem[Bressan et al.(2012)]{bressan12}Bressan, A., Marigo, P., Girardi, L., 2012,  MNRAS, 427, 127B
\bibitem[Butler et al.(1996)]{butler:1996} Butler, R. P., Marcy, G. W.,
    Williams, E., McCarthy, C., Dosanjh, P., \& Vogt, S. S.  1996,
    \pasp, 108, 500
\bibitem[Chabrier et al.(2000)]{chabrier00}Chabrier, G., Baraffe, I., Allard, F., \& Hauschildt, P.: 2000, ApJ, 542, 464
\bibitem[Chen et al.(2015)]{chen15}Chen, Y., Bressan, A., Girardi, L., et al. 2015, MNRAS, 452, 1068
\bibitem[Currie(2009)]{currie09}Currie, T., 2009, ApJ, 694L, 171C
\bibitem[Cutri et al.(2003)]{cutri03}Cutri, R. M., Skrutskie, M. F., van Dyk, S., et al. 2003, VizieR Online Data Catalog, 2246, 0
\bibitem[Crepp et al.(2012)]{crepp12}Crepp, J. R., Johnson, J. A., Howard, A. W., et al. 2012, ApJ, 761, 39C
\bibitem[Crepp et al.(2013a)]{crepp13a}Crepp, J. R., Johnson, J. A., Howard, A. W., et al. 2013a,  ApJ, 771, 46C
\bibitem[Crepp et al.(2013b)]{crepp13b}Crepp, J. R., Johnson, J. A., Howard, A. W., et al. 2013b,  ApJ, 774, 1C
\bibitem[Crepp et al.(2014)]{crepp14}Crepp, J. R., Johnson, J. A., Howard, A. W., et al. 2014,  ApJ, 781, 29C
\bibitem[da Silva et al.(2011)]{dasilva11}da Silva, R., Milone, A. C., Reddy, B. E. 2011,  A\&A, 526A, 71D
\bibitem[Duch\'ene \& Kraus(2013)]{duchene13}Duch\'ene, G., \& Kraus, A. 2013, ARA\&A, 51, 269D
\bibitem[Durisen et al.(2007)]{durisen07}Durisen, R. H., Boss, A. P., Mayer, L., et al. 2007, in Protostars and Planets V, ed. B. Reipurth, D. Jewitt, \& K. Keil (Tucson, AZ: Univ. Arizona Press), 607
\bibitem[Egner et al.(2010)]{egner10}Egner, S., Ikeda, Y., Watanabe, M., et al. 2010,  SPIE, 7736E, 4VE
\bibitem[Ford(2005)]{ford:2005} Ford, E.B. 2005, \aj, 129, 1706
\bibitem[Ford (2006)]{ford:2006} Ford, E. B. 2006,
    New Horizons in Astronomy: Frank N. Bash Symposium 352, 15.
\bibitem[Ford \& Gregory(2007)]{ford:2007} Ford E. B., Gregory P. C., 2007,
in Babu G. J., Feigelson E. D., eds, ASP Conf. Ser. Vol. 371, Statistical Challenges
in Modern Astronomy IV. Astron. Soc. Pac., San Francisco, p. 189 
\bibitem[Gelman \& Rubbin(1992)]{gelman:1992} Gelman, A. \& Rubin, D.B.
1992, Statistical Science, 7, 457
\bibitem[Gregory(2005)]{gregory:2005} Gregory, P.C. 2005, \apj, 631, 1198
\bibitem[Hauschildt et al.(1999a)]{hauschildt99a}Hauschildt, P. H., Allard, F., \& Baron, E., 1999a, ApJ, 512, 377
\bibitem[Hauschildt et al.(1999b)]{hauschildt99b}Hauschildt, P. H., Allard, F., Ferguson, J., Baron, E., \& Alexander, D. R.: 1999b, ApJ, 525, 871
\bibitem[Holman et al.(1997)]{holman97}Holman, M., Touma, J., \& Tremaine, S. 1997, Nature 386, 254
\bibitem[Howard et al.(2010)]{Howard10}
Howard, A. W., Johnson, J. A., Marcy, G. W., et al. 2010, ApJ, 721, 1467H
\bibitem[Hayano et al.(2008)]{hayano08}Hayano, Y., et al. 2008, SPIE, 7015E, 10H
\bibitem[Izumiura (1999)]{izumiura99}Izumiura, H. 1999, in Proc. 4th East Asian Meeting on Astronomy,
ed. P. S. Chen (Kunming: Yunnan Observatory), 77
\bibitem[Janson et al.(2013)]{janson13}Janson, M., Lafreni\`ere, D., Jayawardhana, R., et al. 2013, ApJ, 773, 170
\bibitem[Kambe et al.(2002)]{kambe02}Kambe, E., Sato, B., Takeda, Y., et al. 2002, PASJ, 54, 865
\bibitem[Kane et al.(2010)]{kane10}Kane, S. R., Reffert, S., Henry, G. W., et al. 2010,  ApJ, 720, 1644K
\bibitem[Kane et al.(2014)]{kane14}Kane, S. R., Howell, S. B., Horch, E. P., 2014, ApJ, 785, 93K
\bibitem[Kley \& Nelson(2012)]{kley12}Kley, W., \& Nelson, R. P. 2012, ARA\&A, 50, 211
\bibitem[Knutson et al.(2014)]{knutson14}Knutson, H. A., Fulton, B. J., Montet, B. T., et al. 2014, ApJ, 785, 126
\bibitem[Kuzuhara et al.(2013)]{kuzuhara13}Kuzuhara, M., Tamura, M., Kudo, T., et al. 2013, ApJ, 774, 11
\bibitem[Lafreni\`ere et al.(2007a)]{lafreniere07a}Lafreni\`ere, D., Marois, C., Doyon, R.,  Nadeau, D., \& Artigau, E. 2007, ApJ, 660, 770L
\bibitem[Lafreni\`ere et al.(2007b)]{lafreniere07b}Lafreni\`ere, D., Doyon, R., Marois, C., et al. 2007ApJ, 670, 1367
\bibitem[Lejeune and Schaerer(2001)]{Lejeune2001}Lejeune, T., \& Schaerer, D. 2001, A\&A, 366, 538
\bibitem[Liu et al.(2002)]{liu02}Liu, M. C., Fischer, D. A., Graham, J. R., et al. 2002, ApJ, 571, 519
\bibitem[Marois et al.(2006)]{marois06}Marois, C., Lafreni\`ere, D., Doyon, R., et al. 2006,  ApJ, 641, 556M
\bibitem[Martinache \& Guyon(2009)]{Martinache09}Martinache, F., \& Guyon, O. 2009, Proc. SPIE, 7440, 0
\bibitem[Montet et al.(2014)]{Montet14}	
Montet, B. T., Crepp, J. R.,  Johnson, J. A., Howard, A. W., and Marcy, G. W. 2014, ApJ, 781, 28M
\bibitem[Mugrauer et al.(2014)]{mugrauer14}Mugrauer, M., Ginski, C., Seeliger, M. 2014,  MNRAS, 439, 1063M
\bibitem[Nagasawa et al.(2008)]{nagasawa08}Nagasawa, M., Ida, S., \& Bessho, T. 2008, ApJ, 678, 498
\bibitem[Narita et al.(2010)]{narita10}Narita, N., Kudo, T., Bergfors, C., et al. 2010, PASJ, 62, 779N
\bibitem[Narita et al.(2012)]{narita12}Narita, N., Takahashi, Y. H., Kuzuhara, M., et al. 2012, PASJ, 64L, 7N 
\bibitem[Pollack et al.(1996)]{pollack96}Pollack, J. B., Hubickyj, O., Bodenheimer, P., et al. 1996, Icar,124, 62
\bibitem[Sato et al.(2002)]{sato:2002} Sato, B., Kambe, E.,
    Takeda, Y., Izumiura, H., \& Ando, H.  2002, \pasj, 54, 873
\bibitem[Sato et al.(2003)]{sato03}Sato, B., Ando, H, Kambe, E., et al. 2003,  ApJ, 597L, 157S
\bibitem[Sato et al.(2008)]{sato08}Sato, B., Izumiura, H., Toyota, E., et al. 2008, \pasj, 60, 539
\bibitem[Sato et al.(2012)]{sato12}Sato, B., Omiya, M., Harakawa, H., et al. 2012,  \pasj, 64, 135S
\bibitem[Sato et al.(2013a)]{sato13a}Sato, B., Omiya, M., Wittenmyer, R. A., et al. 2013, ApJ, 762, 9
\bibitem[Sato et al.(2013b)]{sato13b} Sato, B., Omiya, M., Harakawa, H., et al. 2013, \pasj, 65, 85
\bibitem[Suzuki et al.(2010)]{suzuki10}Suzuki, R., Kudo, T., Hashimoto, J., et al. 2010, SPIE, 7735E, 30S
\bibitem[Takahashi et al.(2013)]{takahashi13}Takahashi, Y. H., Narita, N., Hirano, T., et al. 2013 arXiv: 1309.2559T
\bibitem[Takeda et al.(2008)]{takeda08}Takeda, Y., Sato, B., Murata, D. 2008,  PASJ, 60, 781T
\bibitem[Tamura(2009)]{tamura09}Tamura, M. 2009, AIPC, 1158, 11T
\bibitem[Torres(1999)]{torres99}Torres, G. 1999, \pasp, 111, 169
\bibitem[\protect\citeauthoryear{Valenti et~al.} {1995}]{val:95}
    Valenti, J.~A., Butler, R.~P. \& Marcy, G.~W. 1995,
    \newblock { PASP, } {107}, 966.
\bibitem[van Dokkum(2001)]{vandokkum01}van Dokkum, Pieter G. 2001PASP..113.1420
\bibitem[van Leeuwen (2007)]{van07}van Leeuwen, F. 2007, \aap, 474, 653
\bibitem[Vigan et al.(2012)]{vigan12}Vigan, A., Patience, J., Marois, C., et al. 2012, A\&A, 544, 9
\bibitem[Wang et al.(2014)]{wang14}Wang, L., Sato, B., Omiya, M., 2014,  PASJ, 66, 118
\bibitem[Wu \& Murray(2003)]{wu03}Wu, Y. \& Murray, N. 2003, ApJ, 589, 605
\bibitem[Wu \& Lithwick(2010)]{wu10}Wu, Y., \& Lithwick, Y. 2010, ApJ, 735, 109
\bibitem[Zechmeister et al.(2008)]{zechmeister08}Zechmeister, M., Reffert, S., Hatzes, A. P., et al. 2008, 
 A\&A, 491, 531Z
\end{thebibliography}
\end{document}